\documentclass[%
 reprint,
 amsmath,amssymb,
 aps,
]{revtex4-2}

\usepackage{graphicx}% Include figure files
\usepackage{dcolumn}% Align table columns on decimal point
\usepackage{bm}% bold math
\usepackage{braket}
\usepackage{xcolor}
\usepackage{afterpage} % flush floating figures
\newcommand{\mean}[1]{\mathcal{M}\left[#1\right]}
\newcommand{\fii}{\varphi}
\newcommand{\abs}[1]{\left\vert #1\right\vert}
\newcommand{\ip}[2]{\langle#1\vert#2\rangle}
\newcommand\numberthis{\addtocounter{equation}{1}\tag{\theequation}}

\def\ketbra#1#2{\ket{#1}\bra{#2}}

\begin{document}

\title{Phonon-Induced Effects in Quantum Dot Absorption and Resonance Fluorescence with Hierarchy of Pure States}% Force line breaks with \\

\author{Sebastian Toivonen}
\affiliation{Department of Physics and Astronomy, University of Turku, Turku, Finland}
\author{Kimmo Luoma}
\email{ktluom@utu.fi}
\affiliation{Department of Physics and Astronomy, University of Turku, Turku, Finland}

\date{\today}% It is always \today, today,

\begin{abstract}
We investigate a quantum dot (QD) system coupled to a vibrational environment with a super-Ohmic spectral density and weakly to a leaky cavity mode, a model relevant for semiconductor-based single-photon sources. The phonon coupling induces dephasing and broadens the absorption and emission line shapes, while the weakly coupled cavity mode leads to effective driving of the QD.%modifies the emission rate via Purcell enhancement. 
To capture non-Markovian effects, we use non-Markovian Quantum State Diffusion and its hierarchical extension the Hierarchy of Pure States to compute multitime correlation functions underlying absorption and resonance fluorescence spectra. We present numerical results for the absorption spectra at strong phonon coupling and finite temperature, as well as for resonance fluorescence spectra at varying phonon coupling strengths and temperatures, and analyse the visibility of the resonance fluorescence spectra to provide insights into how phonon coupling and thermal effects influence the spectral features.
\end{abstract}

%\tableofcontents

\maketitle
\section{Introduction}
In this work, we investigate a quantum dot (QD) system that features strong coupling to its vibrational environment and weak coupling to a leaky cavity mode. This model serves as a simplified representation of a semiconductor quantum dot-based single-photon source, for example. Despite its simplicity, the model captures many essential features of the real system: phonon coupling causes dephasing and broadens the absorption lineshapes, while the weakly coupled leaky cavity leads to effective classical driving of the QD. 
%Remove Purcell enhancement as we do not have any radiative damping in the model (at least written here).
%system emission rate via Purcell enhancement. 
Using this model, many of the features observed in recent experiments \cite{Jordan2024, Portalupi2015, Phillips2024, Siampour2023, Daveau2017, Giesz2013, Abdullah2019, Liu2018} can be at least qualitatively understood. A solid understanding of this model is crucial for improving quantum dot-based single-photon sources, where the quality of emitted single-photons is limited by vibrational coupling \cite{Denning2020_1, Senellart2017, Dreesen2018, Bozzio2022, Buckley2012, Gustin2018}.

Many important experimental observables, such as emission spectra, can be computed from multitime expectation values. Systems described by the Gorini-Kossakowski-Sudarshan-Lindblad (GKSL) equation \cite{Lindblad1976, Gorini1976} often exhibit the useful property that these multitime observables satisfy the same equation of motion as the density matrix -- this is the essence of the Quantum Regression Theorem (QRT) \cite{Carmichael1993, Gardiner2004, Breuer2007}. The validity and limitations of the QRT have been extensively investigated in recent years, leading to an understanding of when the theorem applies and when it breaks down \cite{Guarnieri2014, Cosacchi2021, Khan2024, Khan2022, Lonigro2022, Panyukov2024, Boedecker2012, Talkner1986, Alonso2005}.

Clearly, computation of multitime expectation values for systems beyond GKSL paradigm is an important topic as the large body of research in recent years suggests \cite{Alonso2005, Lonigro2022, Tarasov2021, Dowling2023, Burgarth2021, Zambon2024, Smirne2022, Campaioli2024, Starthearn2018, Shurikant2023}. In this work, we contribute to this effort. We compute multitime observables using Non-Markovian Quantum State Diffusion (NMQSD) and it's hierarchical extension Hierarchy of Pure States (HOPS). %
We focus in particular on the case where the coupling to the non-Markovian environment is described by a super-Ohmic spectral density, which is typical
for quantum dots. We investigate how the the presence of a weakly coupled cavity mode is modifying the bare absorption spectrum and resonance fluorescence spectrum of the quantum dot. 

The outline of the article is as follows: In Sec. \ref{section_model}, we introduce the model describing the quantum dot, its coupling to the phonon environment and the cavity. This section also includes a brief review of the NMQSD and the HOPS method, which form the basis of our calculations. Sec. \ref{section_results} presents the numerical results, including the calculation of the QD linear absorption spectrum under strong phonon coupling and at finite temperatures. We also analyse the correlation functions of the independent boson model and the resonance fluorescence spectra. We examine how their visibility varies with coupling strength and temperature. Finally, Sec. \ref{section_conclusions} provides a discussion and conclusions. The appendices detail the adiabatic elimination of the cavity and additional details on the  numerical results related to the resonance fluorescence spectra. In this work we set $\hbar = 1$.

\section{Model} \label{section_model}
In this work we consider a driven semiconductor QD that is placed in a cavity. The exciton, electron-hole pair, is coupled both to the cavity modes and the surrounding lattice vibrations. The QD is modeled as a two-level system with states $\ket{g}$ and $\ket{X}$ corresponding to the ground and excited states, respectively, and the system Hamiltonian is given by $H_\text{S} = \omega_\text{X} \ketbra{X}{X}$. The lattice vibrations, or phonons, are modeled as quantum harmonic oscillators. The free field Hamiltonian of the phonons is given (in the harmonic approximation) by $H_\text{Ph} = \sum_\lambda \omega_\lambda a_\lambda^\dagger a_\lambda$, where $a^\dagger_\lambda$ and $a_\lambda$ are creation and annihilation operators of the mode $\omega_\lambda$. We assume a linear coupling between the QD and the environment. The interaction Hamiltonian is in this case given by
\begin{align*}
H_\text{I,Ph} = \ketbra{X}{X} \sum_\lambda g_\lambda (a_\lambda^\dagger + a_\lambda).
\end{align*}
The coupling strength of each phonon mode is encoded in the coefficients $g_\lambda \in \mathbb{R}$. To measure the coupling strength in the continuum limit, the exciton-phonon spectral density is defined by
\begin{align}
\label{spectral_density_definition}
J_\text{Ph}(\omega) = \sum_\lambda |g_\lambda|^2 \delta(\omega - \omega_\lambda).
\end{align}
We assume that the dominant exciton-phonon coupling mechanism is the deformation potential \cite{Ziman2001}. As such the spectral density has the explicit form
\begin{align}
J_\text{Ph}(\omega) = A \omega^3 e^{-\frac{\omega^2}{\xi^2}}
\end{align}
with $\xi \in \mathbb{R}$ being the cut-off frequency and $A \in \mathbb{R}$ denotes the coupling strength. The influence of the lattice vibrations on the QD is encoded in the bath-correlation function (BCF) 
\begin{align}
    \alpha(t-s) = \mathrm{tr}\left\{\rho_\text{Ph}Q(t)Q(s)\right\},
\end{align}
where $Q(t) = \sum_\lambda g_\lambda\left(a_\lambda e^{-i\omega_\lambda t}+a_\lambda^\dagger e^{i\omega_\lambda t}\right)$. In the finite temperature case 
$\rho_\text{Ph}=\frac{1}{Z}e^{-\beta H_\text{Ph}}$ and 
\begin{align*}
\alpha(\tau) = \int^\infty_0 \text{d} \omega J_\text{Ph}(\omega) \left( \text{coth}\left( \frac{\beta \omega}{2}\right) \cos \omega \tau - i \sin \omega \tau \right),
\end{align*}
with $\tau = t - s$ and the inverse temperature $\beta = 1 / T$. At zero temperature this reduces simply to the Fourier integral
\begin{align*}
\alpha(\tau) = \int^\infty_0 \text{d} \omega J_\text{Ph}(\omega) e^{-i \omega \tau}.
\end{align*}

The cavity and it's associated coupling is incorporated in a quantized manner while the external driving is treated as a classical field. We assume that a single cavity mode $\omega_c$ is present. The Hamiltonian of the driven cavity is given by
\begin{align*}
H_{C + D} = &\omega_c b^\dagger b + g (\ketbra{X}{g} b + \ketbra{g}{X} b^\dagger) \\
&+ f(t) b^\dagger + f^*(t) b,
\end{align*}
where $b^\dagger$ and $b$ are the field creation and annihilation operators and $f(t)$ is an arbitrary drive and $g$ describes the coupling strength between the cavity and the exciton. We eliminate the cavity operators by adiabatic elimination. Detailed calculation is presented in the appendix. This results in the Hamiltonian
\begin{align*}
\tilde{H}_{C + D} = &\omega_c |\langle b \rangle|^2 + g (\ketbra{X}{g} \langle b \rangle + \ketbra{g}{X}\langle b^\dagger \rangle) \\
&+ f(t) \langle b^\dagger \rangle + f^*(t) \langle b \rangle,
\end{align*}
where $\langle b \rangle$ is given by
\begin{align*}
\langle b \rangle = \frac{f(t)}{-\omega_c + i \frac{\kappa}{2}},
\end{align*}
where $\kappa>0$ is the cavity damping rate.
In the case of monochromatic continuous wave driving, we set $f(t) = \Omega' \cos \omega_D t$. Applying the rotating wave approximation and moving to the frame rotating at the drive frequency $\omega_D$ results in the Hamiltonian
\begin{align}
\tilde{H}_{C+D} = \tilde{\Omega}(\omega_c \sigma_x + \frac{\kappa}{2} \sigma_y),
\end{align}
where $\tilde{\Omega} = \frac{-g \Omega'}{2 \omega_c^2 + \frac{\kappa^2}{2}}$. The terms proportional to the identity operator have been omitted since they lead to a global phase only.%will not have an effect on the corresponding master equation. 
 We also define $\Omega = \Omega' \omega_c$.

\subsection{Hierarchy of pure states}
To solve the dynamics of the QD system, we will use the hierarchy of pure states (HOPS) method \cite{Suess2014}. HOPS is a numerically exact method developed to solve the equations emerging from non-Markovian quantum state diffusion (NMQSD) \cite{Diosi1997, Diosi1998}. The NMQSD approach aims to solve the dynamics of both the system and its environment for a given open quantum system. The environment is assumed to consist of quantum harmonic oscillators and the total Hamiltonian is assumed to be of the form
\begin{align*}
H = H_S + H_E + H_I,
\end{align*}
where $H_S$ is the system Hamiltonian, $H_E = \sum_\lambda \omega_\lambda a^\dagger a$ is the environment Hamiltonian and 
\begin{align*}
H_I = \sum_\lambda (g_\lambda L a^\dagger + g^*_\lambda L^\dagger a)
\end{align*}
is the system-environment interaction Hamiltonian. Here $L$ is the coupling operator, which is an operator of the system. At zero temperature the initial total state is assumed to be factorized $\ket{\Psi_0} = \ket{\psi}\ket{\mathbf{0}}$. The reduced density operator is given by the ensemble average
\begin{align}
\label{ensemble_average}
\rho_S(t) = \text{tr}_E \{ \ketbra{\Psi_t}{\Psi_t} \} = \mathcal{M}[\ketbra{\psi_t(z^*)}{\psi_t(z)}].
\end{align}
The average is taken over the stochastic trajectories $\ket{\psi_t(z^*)}$ and these obey the linear NMQSD equation
\begin{align*}
\partial_t \ket{\psi_t(z^*)} =& -iH_S \ket{\psi_t(z^*)} + z_t^* L \ket{\psi_t(z^*)} \\
\numberthis \label{NMQSD_equation} &- L^\dagger \int^t_0 \text{d}s \alpha(t - s) \frac{\delta}{\delta z_s^*} \ket{\psi_t(z^*)},
\end{align*}
where $z_t^*$ is a Gaussian stochastic process with zero mean and correlations
\begin{align*}
\mathcal{M}[z_t z_s] = 0,~ \mathcal{M}[z_t z_s^*] = \alpha(t - s).
\end{align*}
The norm of the state vector $\ket{\psi_t(z^*)}$ is not preserved and as such only a few rare  trajectories dominate the ensemble average (\ref{ensemble_average}). To ensure more rapid convergence, it is desirable to take the ensemble average over normalized trajectories such that each trajectory contributes with equal weight to the average
\begin{align*}
\rho_S(t) = \mathcal{M} \left[ \frac{\ketbra{\psi_t(\tilde{z}^*)}{\psi_t(\tilde{z})}}{||\psi_t(\tilde{z}^*)||^2} \right].
\end{align*}
This requires some modifications to the linear NMQSD equation \cite{Diosi1998}. The noise process is replaced by a shifted noise process
\begin{align}
\label{shifted_noise_process}
\tilde{z}_t^* = z_t^* + \int^t_0 \text{d} s \alpha^*(t - s) \langle L^\dagger \rangle_s
\end{align}
and the operator $L^\dagger$ is replaced with $L^\dagger - \langle L^\dagger \rangle_t$ where the expectation value is calculated using the normalized state vector
\begin{align*}
\langle L^\dagger \rangle_t = \frac{\braket{\psi_t(\tilde{z}) | L^\dagger | \psi_t(\tilde{z}^*)}}{\braket{\psi_t(\tilde{z}) | \psi_t(\tilde{z}^*)}}.
\end{align*}
These replacements lead to the non-linear NMQSD equation
\begin{align*}
\partial_t \ket{\psi_t(\tilde{z}^*)} =& -iH_S \ket{\psi_t(\tilde{z}^*)} + \tilde{z}^*_t L \ket{\psi_t(\tilde{z}^*)} \\
&- (L^\dagger - \langle L^\dagger \rangle_t ) \int^t_0 \text{d}s \alpha(t - s) \frac{\delta}{\delta \tilde{z}_s^*} \ket{\psi_t(\tilde{z}^*)}.
\end{align*}

Finite temperature can be incorporated by adding a stochastic potential to the Hamiltonian \cite{Hartmann2017, Goetsch1996}
\begin{align*}
H_S \rightarrow H_S + V(t),
\end{align*}
where $V(t) = \eta^*_t L + \eta_t L^\dagger$, with the Gaussian stochastic process $\eta_t$ obeying the statistics
\begin{align*}
\mathcal{M}[\eta_t] = \mathcal{M}[\eta_t \eta_s] = 0,~\mathcal{M}[\eta_t \eta_s^*] = \sum_\lambda \bar{n}_\lambda |g_\lambda|^2 e^{-i\omega_\lambda(t-s)},
\end{align*}
where $\bar{n}_\lambda = (e^{\beta \omega_\lambda} - 1)^{-1}$ is the Bose-Einstein distribution for the environmental mode $\omega_\lambda$.

The difficulty in solving the NMQSD equations arise from the functional derivatives. The HOPS method was developed to solve the NMQSD equations in a numerically exact manner. Essentially the functional derivative is turned into a hierarchy of auxiliary states by assuming that the BCF takes the form of a sum of exponentials
\begin{align*}
\alpha(\tau) = \sum^K_{\mu = 1} \mathcal{G}_\mu e^{-W_\mu \tau},
\end{align*}
with $\mathcal{G}_\mu, W_\mu \in \mathbb{C}$. The linear HOPS equation corresponding to the linear NMQSD equation (\ref{NMQSD_equation}) is given as \cite{Suess2014}
\begin{align*}
\partial_t \ket{\psi_t^\mathbf{k}(z^*)} =& \left(-iH_S - \sum_{\mu = 1}^K k_\mu W_\mu + z^*_t L \right) \ket{\psi_t^\mathbf{k}(z^*)} \\
\numberthis \label{linear_HOPS} &+ L \sum_{\mu = 1}^K k_\mu \mathcal{G}_\mu \ket{\psi_t^{\mathbf{k} - \mathbf{e}_\mu}(z^*)} \\
&- L^\dagger \sum_{\mu = 1}^K \ket{\psi_t^{\mathbf{k} + \mathbf{e}_\mu}(z^*)},
\end{align*}
where $\mathbf{k} = (k_1, k_2, ..., k_K)$ with $k_\mu \in \mathbb{N}, k_\mu \geq 0$. Note that this is a system, or hierarchy, of coupled stochastic differential equations. Solving this hierarchy corresponds to solving the NMQSD equation for $\ket{\psi_t(z^*)} = \ket{\psi^\mathbf{0}_t(z^*)}$. The remaining states $\ket{\psi_t^\mathbf{k}(z^*)}$ are 'auxiliary states,' illustrating how the finite memory time of the environment influences the time evolution of the QD. For practical calculations, the hierarchy is truncated based on a specified condition, with the last level set as a terminator. In this work, we employ a triangular truncation condition where auxiliary states for which $|\mathbf{k}| = \sum_\mu k_\mu > \mathcal{K}$, for some predetermined $\mathcal{K}$, are set to zero. More sophisticated truncation methods have been investigated as well \cite{Hartmann2021}. The corresponding HOPS for the non-linear NMQSD equation is
\begin{align*}
\partial_t \ket{\psi_t^\mathbf{k}(\tilde{z}^*)} =& \left(-iH_S - \sum_{\mu = 1}^K k_\mu W_\mu + \tilde{z}^*_t L \right) \ket{\psi_t^\mathbf{k}(\tilde{z}^*)} \\
&+ L \sum_{\mu = 1}^K k_\mu \mathcal{G}_\mu \ket{\psi_t^{\mathbf{k} - \mathbf{e}_\mu}(\tilde{z}^*)} \\
\numberthis \label{non_linear_HOPS} &- (L^\dagger - \langle L^\dagger \rangle_t) \sum_{\mu = 1}^K \ket{\psi_t^{\mathbf{k} + \mathbf{e}_\mu}(\tilde{z}^*)},
\end{align*}
where $\tilde{z}_t^*$ is the shifted noise process (\ref{shifted_noise_process}) and $\langle L^\dagger \rangle_t$ is given by the zeroth level state
\begin{align*}
\langle L^\dagger \rangle_t = \frac{\braket{\psi^\mathbf{0}_t(\tilde{z})|L^\dagger|\psi^\mathbf{0}_t(\tilde{z}^*)}}{\braket{\psi^\mathbf{0}_t(\tilde{z})|\psi^\mathbf{0}_t(\tilde{z}^*)}}.
\end{align*}
It is worth noting that the zero-temperature BCF is used for the exponential expansion, and the temperature is incorporated through the stochastic potential without modification to the hierarchy structure.

\section{Results} \label{section_results}
\subsection{Independent boson model}
\begin{figure}
\begin{center}
\includegraphics[scale=0.55]{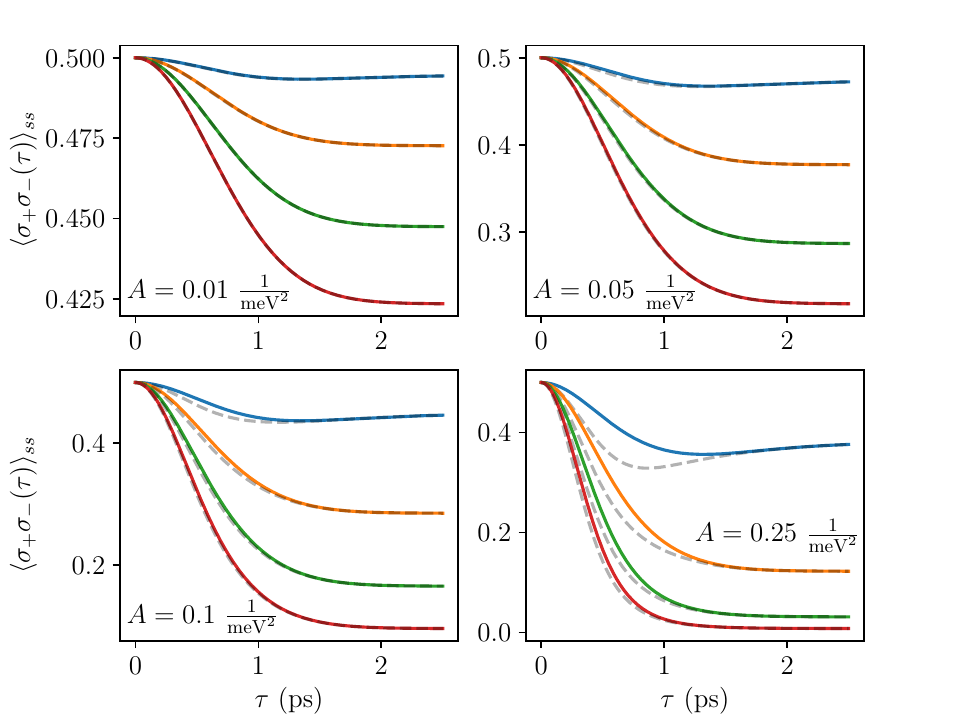}
\caption{Correlation function $\langle \sigma_+ \sigma_-(\tau) \rangle_\text{ss}$ for the independent boson model at different temperatures with fixed phonon coupling strengths for the factorized steady state (solid) and exact steady state (dashed). The temperatures are from top to bottom in each panel: $0$ K, $25$ K, $50$ K and $75$ K.}\label{IBM_fixed_alpha_correlation_functions}
\end{center}
\end{figure}
The phonons of the environment are modeled as independent bosons. This model, with only the QD and the phonon environment present, is known as the independent boson model (IBM) or pure dephasing model \cite{Kok2010, Mahan2000, Krummheuer2002, Hohenster2007}. The IBM admits an exact solution, from which we have for the reduced density matrix \cite{Toivonen2023}
\begin{align*}
\rho_S(t) =&~ \rho_{gg} \ketbra{g}{g} + e^{i \omega'_Xt} e^{-\Gamma(t)} \rho_{gX} \ketbra{g}{X} \\&+ e^{-i \omega'_Xt} e^{-\Gamma^*(t)} \rho_{Xg} \ketbra{X}{g} + \rho_{XX} \ketbra{X}{X},
\end{align*}
where $\omega'_X = \omega_X - \int \text{d}\omega \frac{J(\omega)}{\omega}$ is the phonon-shifted transition energy of the QD and $\Gamma(t)$ is the decoherence function
\begin{align*}
\Gamma(t) = \int \text{d} \omega \frac{J(\omega)}{\omega^2} \left( \coth \left( \frac{\beta \omega}{2} \right) (1 - \cos \omega t) - i \sin \omega t \right).
\end{align*}
It is often convenient to approximate the steady state of the system and the environment as the uncorrelated state.  To see the validity of this approximation, we calculate the exact correlation function
\begin{align*}
\langle \sigma_+ \sigma_-(\tau) \rangle_{ss} = \text{tr} \{ \sigma_+ \sigma_-(\tau) \rho_{ss} \}
\end{align*}
for both the approximate case and exact case in the frame rotating at $\omega'_X$. For the approximate case, we find
\begin{align*}
\langle \sigma_+ \sigma_-(\tau) \rangle^{\text{approx}}_{ss} &= 
\text{tr}\{(\rho_S(\infty) \otimes \rho_E^\beta)\sigma_+\sigma_-(\tau)\}\\
&=\rho_{XX} \exp{\left( \Gamma(\tau) \right)}.
\end{align*}
and for the exact case 
\begin{align*}
\langle \sigma_+ \sigma_-(\tau) \rangle_{ss} = \rho_{XX} \exp{\left( \Gamma(\tau) + 2i \int \text{d}\omega \frac{J(\omega)}{\omega^2} \sin \omega \tau \right)}.
\end{align*}

\begin{figure}
\begin{center}
\includegraphics[scale=0.55]{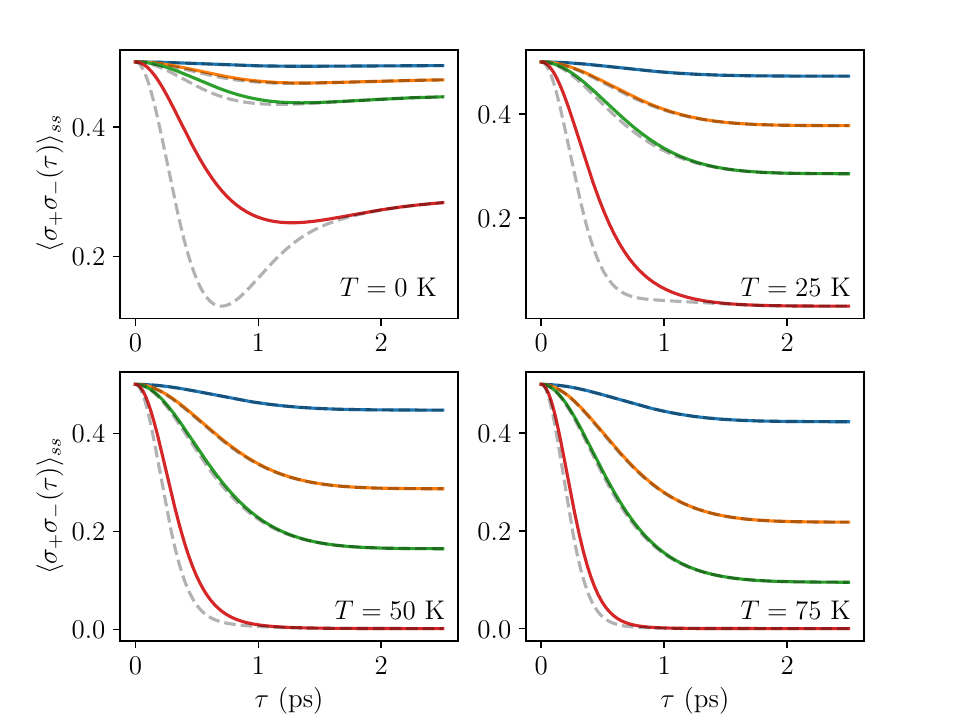}
\caption{Correlation function $\langle \sigma_+ \sigma_-(\tau) \rangle_\text{ss}$ for the independent boson model at different phonon coupling strengths with fixed temperatures for the factorized steady state (solid) and exact steady state (dashed). The phonon coupling strengths are from top to bottom in each panel: $0.01$, $0.05$, $0.1$ and $0.5$ in units $\frac{1}{\mathrm{meV}^2}$.}\label{IBM_fixed_temperature_correlations_functions}
\end{center}
\end{figure}
In figures \ref{IBM_fixed_alpha_correlation_functions} and \ref{IBM_fixed_temperature_correlations_functions}, the real part of the correlation functions for the approximate case (solid lines) and the exact case (dashed lines) are shown for varying temperatures and coupling strengths. At weak coupling strengths, the correlation functions are in good agreement. However, at stronger coupling strengths, temperature starts to play a significant role in the deviation. This behavior is expected. At low temperatures, the environment is predominantly in its ground state, so its influence on the system is minimal unless the coupling strength is large. In the weak coupling regime, system-environment correlations are negligible, making the uncorrelated steady-state a good approximation. At high temperatures, the environment has a large population of thermal excitations, which introduce significant thermal noise. The system is then primarily subject to thermal fluctuations, and the steady state is dominated 
%by classical thermal effects rather than 
by classical thermal fluctuations rather than 
(quantum) correlations between the system and the environment. As a result, even under relatively strong coupling, the system-environment factorization remains a good approximation.

\subsection{Absorption spectra}
\begin{figure}
\begin{center}
\includegraphics[scale=0.5]{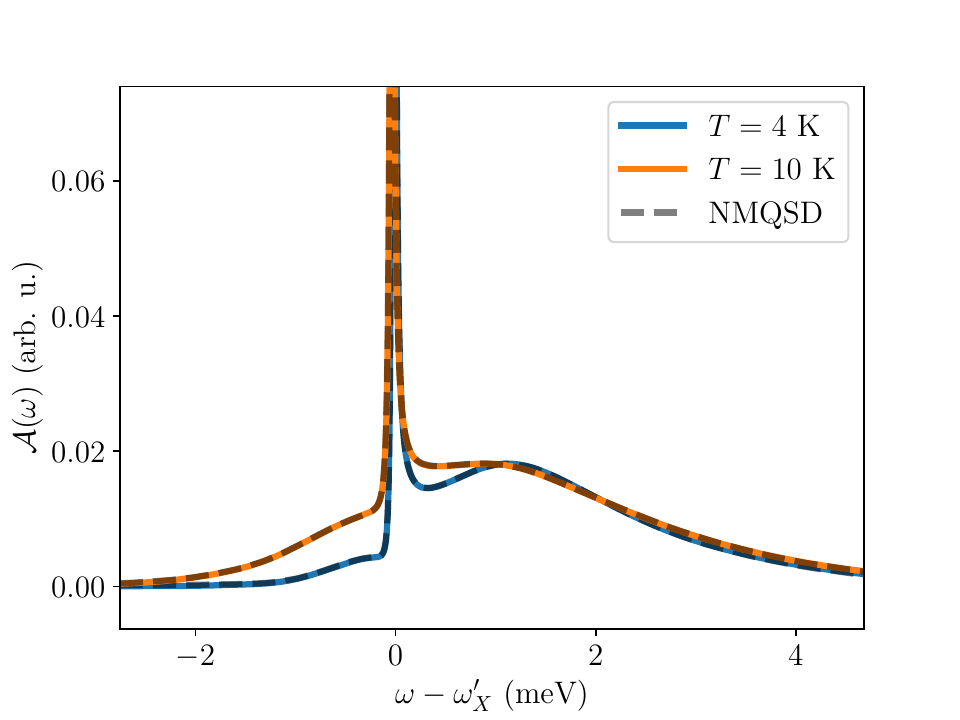}
\caption{Linear absorption spectra of the QD at two different temperatures. A clear asymmetry is visible due to the phonon bath. The colored lines are calculated with HOPS and dashed with NMQSD.}\label{absorption_spectrum}
\end{center}
\end{figure}

\begin{figure}
\begin{center}
\includegraphics[scale=0.5]{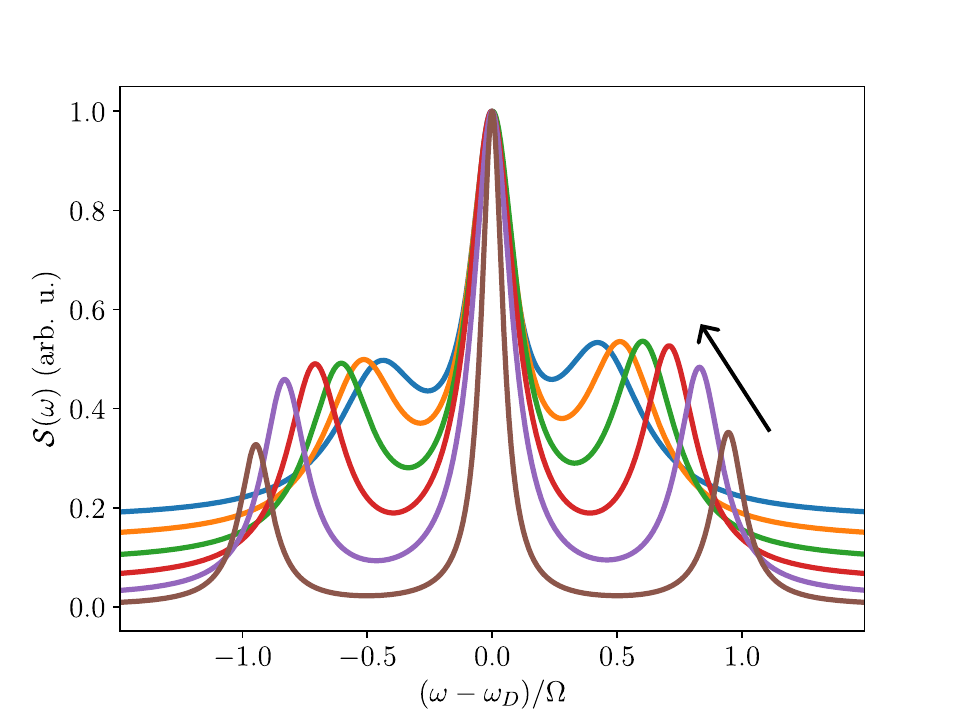}
\caption{Resonance fluorescence spectra of the QD with $A = 0.0175~\frac{1}{\mathrm{meV}^2}$ with temperatures varying from $T = 0~\mathrm{K}$ to $T = 100~\mathrm{K}$ at the top with $20~\mathrm{K}$ increments.}\label{RF_spectra_varying_temperature}
\end{center}
\end{figure}
The linear absorption of the QD can be calculated efficiently with HOPS, by needing to consider only a single trajectory. Since HOPS is numerically exact, and NMQSD doesn't make any assumptions about the strength of the environment coupling, we can simulate spectra in the strong coupling regime very effectively. We consider the Hamiltonian
\begin{align*}
H = H' + H_\text{field}(t),
\end{align*}
where $H'$ is the total Hamiltonian for the system and bath, and $H_\text{field}(t)$ is the Hamiltonian of an electric field
\begin{align*}
H_\text{field}(t) = -\bm{\mu} \cdot \mathbf{E}(t) = -\mu_\text{eff} E(t),
\end{align*}
where $\mu_\text{eff} = \bm{\mu} \cdot \bm{\epsilon}$ is the effective dipole operator with the dipole moment operator $\bm{\mu}$ and polarization $\bm{\epsilon}$. We assume no permanent dipole moment, as such the dipole moment operator is of the form \cite{Kok2010}
\begin{align*}
\bm{\mu} = \tilde{\bm{\mu}} (\ketbra{g}{X} + \ketbra{X}{g}),
\end{align*}
where $\tilde{\bm{\mu}}$ two-level transition dipole moment, which is assumed to be real. The linear absorption spectrum is given by \cite{Kuhn2003, Roden2011}
\begin{align*}
\mathcal{A}(\omega) = \text{Re} \int_0^\infty \text{d} t e^{i\omega t} M(t),
\end{align*}
with the dipole-dipole correlation function
\begin{align*}
M(t) &= \text{tr} \{ \varrho_\text{eq} [\mu_\text{eff}(t), \mu_\text{eff}] \} \\
&= \text{tr}_\text{S} \{\mu_\text{eff} \text{tr}_\text{E}\{ \mathcal{U}_t [\mu_\text{eff}, \varrho_\text{eq}]\mathcal{U}^\dagger_t \}\} \\
&= \text{tr}_\text{S}\{\mu_\text{eff} \kappa(t)\}.
\end{align*}
The equilibrium state is assumed to be factorized $\varrho_\text{eq} = \ketbra{g}{g} \otimes \rho_\beta$. When the cavity is absent, we can calculate the dipole-dipole correlation function by only considering a single trajectory \cite{Roden2011}
\begin{align*}
M(t) \propto \braket{X|\psi_t(z^* = 0)},
\end{align*}
where the stochastic process is set to zero.

Figure \ref{absorption_spectrum} presents the linear absorption spectra of the QD for two different temperatures. Near the QD transition frequency, the spectra are dominated by the phonon spectral density, which enables phonon-assisted photon absorption. With the phonon coupling, a clear asymmetry is observed, which is caused by the differing probabilities of phonon absorption and emission \cite{Roy2012}. These results are consistent with those reported in Ref. \cite{Roy2012}. Without phonon coupling, the spectra would be symmetric Lorentzian peaks and the widths would be determined by radiative broadening. 

\begin{figure}
\begin{center}
\includegraphics[scale=0.5]{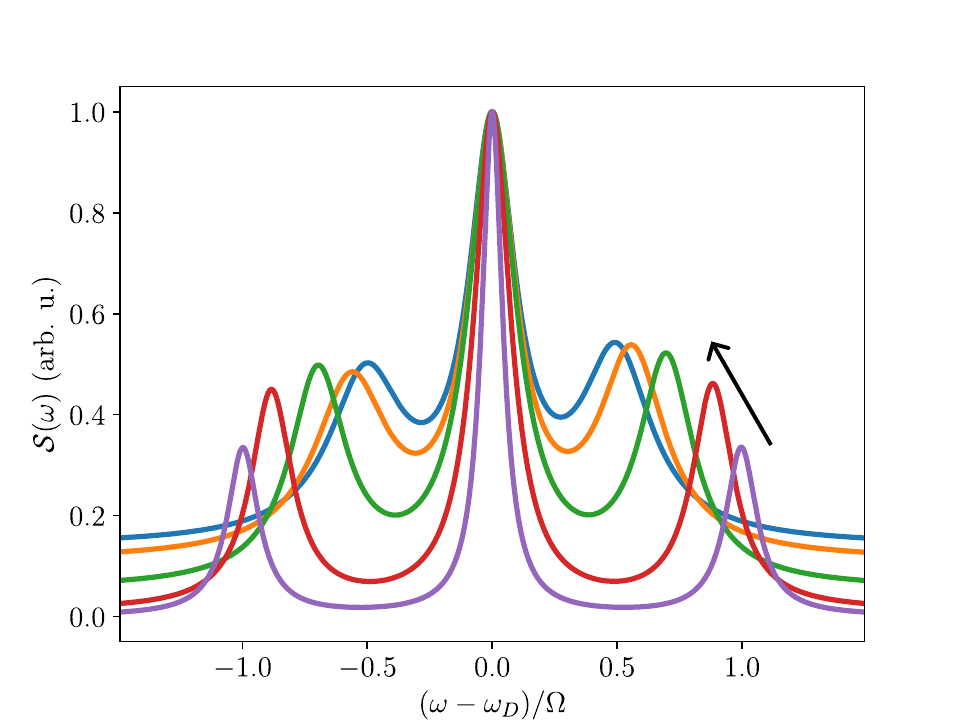}
\caption{Resonance fluorescence spectra of the QD at different coupling strengths at $70~\mathrm{K}$. The coupling strengths $A$ are from bottom to top $0.0$, $0.0175$, $0.0525$, $0.0875$ and $0.105$ in units $\frac{1}{\mathrm{meV}^2}$.}\label{RF_spectra_varying_coupling_strength}
\end{center}
\end{figure}

\begin{figure*}
\begin{center}
\includegraphics[scale=0.65]{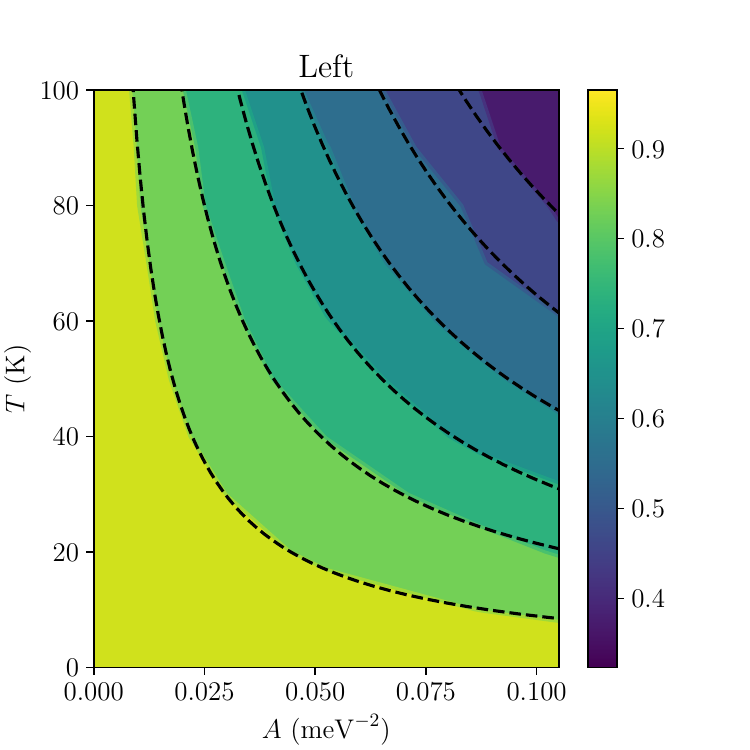}\includegraphics[scale=0.65]{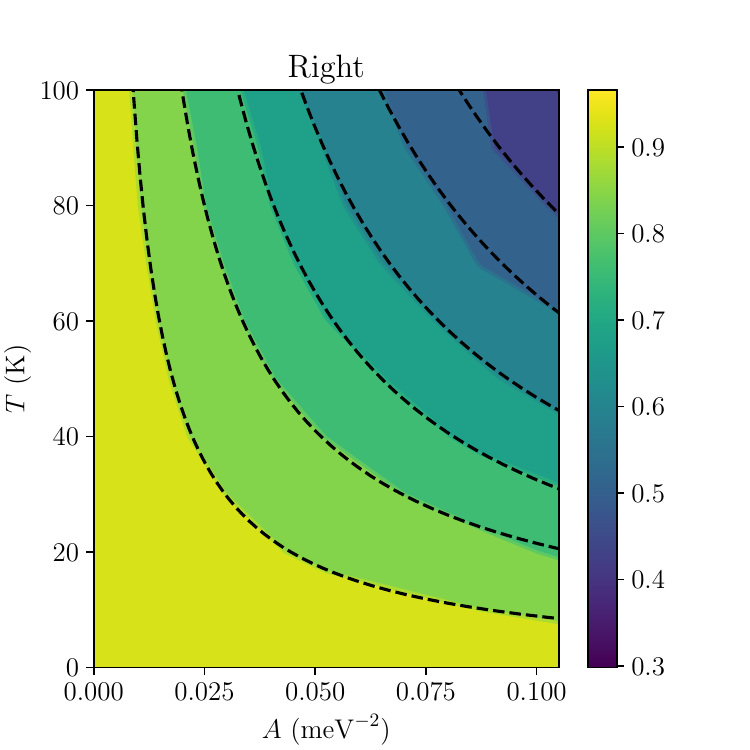}
\caption{Visibilities of the resonance fluorescence spectra side peaks as functions of temperature $T$ and coupling strength $A$. The left figure is for the left peak and right figure for the right peak of the resonance fluorescence spectrum. The dashed lines are the characteristics $T + b / A$ (for some $b$ depending on the curve). Clearly the contour lines follow these characteristics very closely for both of the peaks.}\label{RF_spectra_visibilities}
\end{center}
\end{figure*}

\subsection{Resonance fluorescence spectra}
The spectrum of resonance fluorescence for the QD is given by \cite{Carmichael1993, steck}
\begin{align*}
S(\omega) = \frac{1}{2\pi} \int_{-\infty}^\infty \text{d}\tau e^{i\omega\tau} \langle \sigma_+ \sigma_-(\tau)\rangle_\text{ss},
\end{align*}
where the steady state expectation value is given by $\langle \sigma_+ \sigma_-(\tau)\rangle_\text{ss} = \lim_{t \to \infty} \langle \sigma_+(t) \sigma_-(t + \tau) \rangle$. The spectrum decomposes into two components
\begin{align*}
S(\omega) = S_\text{coh}(\omega) + S_\text{inc}(\omega),
\end{align*}
where $S_\text{coh}(\omega)$ corresponds to coherent scattering and $S_\text{inc}(\omega)$ corresponds to incoherent scattering which are due to quantum fluctuations \cite{Carmichael1993, steck}. These can be explicitly expressed, in the rotating frame of the drive, as
\begin{align*}
S_\text{coh}(\omega) &= \frac{1}{2\pi} \int^\infty_{-\infty} \text{d} \tau  \langle \sigma_+ \rangle_\text{ss} \langle \sigma_- \rangle_\text{ss} e^{-i(\omega - \omega_D) \tau}, \\
S_\text{inc}(\omega) &= \frac{1}{2\pi} \int^\infty_{-\infty} \text{d} \tau \langle \tilde{\sigma}_+ \tilde{\sigma}_-(\tau)\rangle_\text{ss}e^{-i(\omega - \omega_D) \tau},
\end{align*}
where the steady state fluctuations are described by
\begin{align*}
\tilde{\sigma}_\pm(t) = \sigma_\pm(t) - \langle \sigma_\pm \rangle_\text{ss}.
\end{align*}
For the simulations we set $\omega_X=\omega_c=\omega_D=1~\mathrm{eV}$, $\kappa = \omega_c / 10$, $\Omega = 0.175~\mathrm{meV}$, and $g=0.01~\mathrm{meV}$.

The spectra are obtained by fitting three Lorentzian peaks to the numerical data. In figure \ref{RF_spectra_varying_temperature} the resonance fluorescence spectra of the QD is shown at different temperatures. As expected a Mollow triplet structure appears for each of the spectra and the side peaks shift towards the center peak, and the width of the peaks broaden as temperature increases. In figure \ref{RF_spectra_varying_coupling_strength} similar behavior is observed but with fixed temperature and varying phonon-coupling strength.

To evaluate the quality of the spectra, we compute their visibility, as well as their spectral resolutions (presented in the appendix). Visibility is defined as
\begin{align}
\mathcal{V} = \frac{I_\text{max} - I_\text{min}}{I_\text{max} + I_\text{min}},
\end{align}
where $I_\text{max}$ is the height of the central peak, and $I_\text{min}$ is the height of the minimum between the central and sideband peaks. The visiblities of the left and right side peaks are shown in figure \ref{RF_spectra_visibilities} as functions of temperature and phonon coupling strength. Both side peaks exhibit nearly identical visibilities. As seen in figures \ref{RF_spectra_varying_temperature} and \ref{RF_spectra_varying_coupling_strength}, increasing the temperature or phonon coupling strength causes the side peaks to shift closed to the central peak. This shift reduces the visibility of the spectrum, which is reflected in our results. The contours are overlaid with analytic curves $T \propto 1/A$, which align closely with the contour trends, possibly highlighting an underlying relationship. The fluctuations visible in the figure are due to noise in the statisical sampling.

\section{Conclusions and outlook} \label{section_conclusions}
In this work, we have applied the HOPS method to calculate the absorption spectrum of QDs with strong phonon coupling, as well as resonance fluorescence spectra at varying temperatures and phonon coupling strengths. For the absorption spectrum, only a single trajectory was required, even at finite temperature and strong phonon coupling. The calculated absorption spectrum is in good agreement with earlier numerical work. This also indicates that, in this particular case, the approximation of the ground state as a factorized state appears to be valid, as the asymmetry in the spectrum appears. 

To further assess the applicability of the factorized steady-state approximation, we calculated the correlation function $\langle \sigma_+ \sigma_- (\tau) \rangle_{ss}$ for the independent boson model. As expected, at lower temperatures and weak phonon coupling, this approximation works well. However, increasing the phonon coupling strength introduces deviations at short times, after which the correlation functions converge. At higher temperatures, the approximation is less sensitive to increases in phonon coupling strength, maintaining reasonable accuracy. 

The resonance fluorescence spectra were calculated for a range of conditions, from weak phonon coupling and low temperatures to strong phonon coupling and high temperatures. 
%To our best knowledge, this is the first time that the HOPS method has been applied to calculate resonance fluorescence spectra. 
The calculation method is straightforward, as every density matrix propagation can be performed with HOPS without requiring additional computational machinery. From the spectra, we determined their visibility. The contours appear to align closely with curves proportional to $T \propto 1/A$. This suggests that the visibility depends on the characteristics $T + b/A$. Furthermore, this relationship could be used to determine the phonon coupling strength for a given QD system through measurements of resonance fluorescence spectra at different temperatures. 

Nevertheless, certain limitations remain. In this work, we adiabatically eliminated the cavity, which restricts our analysis to weak coupling and leaky cavities. To more fully capture the effects of cavity interactions on QD systems, future work must extend the HOPS method to explicitly include the cavity. Additionally, a well-known limitation of trajectory-based methods is the computational cost of long-time propagation. Depending on how quickly the system reaches its steady state, the required resources can become significant. While the ergodic principle can, in some cases, reduce this cost by allowing the steady state to be obtained through time averaging over a single trajectory \cite{kummerer2004}, this approach was not applicable in our study. Addressing these challenges will be crucial for expanding the applicability of the HOPS method to more complex systems and longer time scales.

In summary, this study advances the application of the HOPS method to strongly coupled QD systems, providing insights into both absorption and resonance fluorescence spectra. It lays the groundwork for future investigations into more complex systems and experimental validations of theoretical predictions.

\begin{acknowledgments}
We thank Konstantinos Daskalakis, Kai Müller, and Walter T. Strunz for illuminating discussions, and CSC for providing the computational resources. S.T. acknowledges the funding from QDOC pilot PhD program.
\end{acknowledgments}

\appendix

\section{Pure state decomposition}
In order to propagate the various states and operators encountered in the main text with HOPS, it is necessary to decompose them into pure states. For a hermitian operator, this decomposition can be obtained through the spectral decomposition
\begin{align*}
\kappa = \sum_\lambda \alpha_\lambda \ketbra{\psi_\lambda}{\psi_\lambda}.
\end{align*}
This also applies to operators that are normal. For other cases we can use the fact that every operator $A$ can be decomposed into a hermitian and an anti-hermitian part
\begin{align*}
A = H + H',
\end{align*}
where $H = \frac{1}{2}(A + A^\dagger)$ and $H' = \frac{1}{2}(A - A^\dagger)$. Since anti-hermitian operators can be expressed as $H' = iK$ where $K$ is a hermitian operator, we can express any operator as the sum
\begin{align*}
A = H + iK,
\end{align*}
where $K = -\frac{i}{2}(A - A^\dagger)$. Since $H$ and $K$ are hermitian, they can be decomposed into pure states by using their respective spectral decompositions
\begin{align*}
A = \sum_\lambda h_\lambda \ketbra{\psi_\lambda^H}{\psi_\lambda^H} + i\sum_\lambda k_\lambda \ketbra{\psi_\lambda^K}{\psi_\lambda^K}.
\end{align*}
Now each of these pure states can be propagated using HOPS
\begin{align*}
\kappa_S(t) &= \text{tr}_E \{ \mathcal{U}_t \kappa \otimes \rho_E \mathcal{U}_t^\dagger \} \\
&= \sum_\lambda \alpha_\lambda \text{tr}_E \{ \mathcal{U}_t \ketbra{\psi_\lambda}{\psi_\lambda} \otimes \rho_E \mathcal{U}_t^\dagger\} \\
&= \sum_\lambda \alpha_\lambda \mathcal{M}[\ketbra{\psi_{\lambda,t}(z^*)}{\psi_{\lambda,t}(z)}],
\end{align*}
where $\ket{\psi_{\lambda,0}} = \ket{\psi_\lambda}$. For the case of a two-level system, this means that we need to propagate four initial states at most. Note that the propagation of each of the state vectors are performed by using the same noise process.

\section{Adiabatic elimination}
\begin{figure*}
\begin{center}
\includegraphics[scale=0.75]{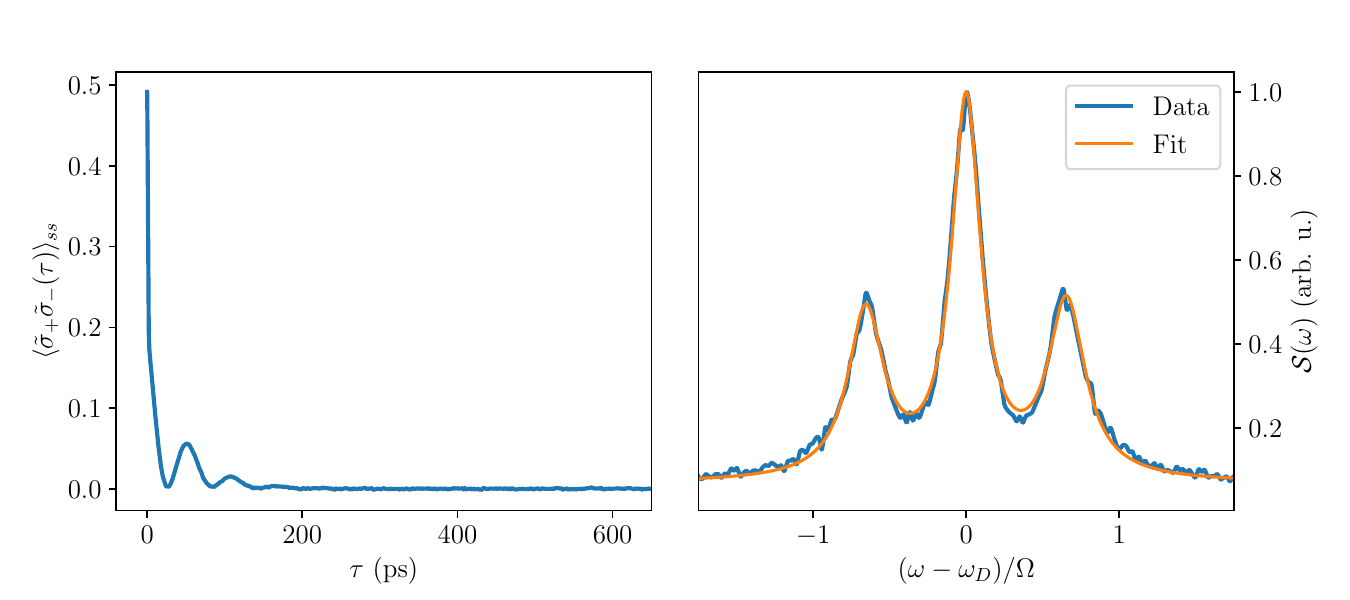}
\caption{The correlation function $\langle \tilde{\sigma}_+ \tilde{\sigma}_- (\tau) \rangle_{ss}$ from the numerical data and its corresponding spectrum. Here the correlation function was obtained with 61000 trajectories.}\label{correlation_function_raw_data_and_fit}
\end{center}
\end{figure*}

The cavity mode can be eliminated in the following way. A model for a continuously monitored cavity mode with heterodyning is given by the following Stratonovich equation
\begin{align}
    \partial_t\ket{\Psi_t} =& -i(H_S+H_C+H_I)\ket{\Psi_t} \notag\\
    &+ \left(a-\langle a \rangle\right)\left(\xi_t^*+\frac{\gamma}{2}\langle a^\dagger\rangle\right)\ket{\Psi_t}\notag\\
    &-\frac{\kappa}{2}\left((a^\dagger -\langle a^\dagger\rangle)a-\langle(a^\dagger - \langle a^\dagger \rangle)a\rangle\right)\ket{\Psi_t},
\end{align}
where $H_S$ is at this point arbitrary, $H_C=\omega_ca^\dagger a+f(t)a^\dagger +f(t)a$ and $H_I = g(La^\dagger +L^\dagger a)$, where $L$ is an arbitrary coupling operator of the system. The Gaussian white noise has zero mean and correlations $\mean{\xi_t\xi_s^*}=\kappa\delta(t-s)$. Other correlations are zero.
Clearly $\ip{\Psi}{\dot\Psi}=0$ and the evolution preserves the norm of the state.
The mean field evolution for the system can be found from a variational principle. We use the following variational ansatz
\begin{align}
    \ket{\Psi_t} = c_t\ket{\fii_t}\ket{z_t},
\end{align}
where $\abs{c(t)}=1$ is a phase factor, $\ket{\fii_t}$ is an arbitrary state for the system and 
$\ket{z}=e^{i\phi}e^{-\frac{1}{2}\abs{z}^2}e^{za^\dagger}\ket{0}$ is a coherent state with a phase factor $\phi$. We thus restrict to a product states where the cavity remains in a coherent state. Using the MacLahlan variational principle we find the following equations of motion
\begin{align}
    \partial_t c_t &= -i c_tE_t,\\
    \partial_t \ket{\fii_t} &= -i(H_{MF,S}-E_t)\ket{\fii_t},\\
    \partial_t\ket{z_t} &= -i(H_{MF,C} - E_t)\ket{z}-\frac{\kappa}{2}(z_ta^\dagger -\abs{z_t})\ket{z},
\end{align}
where $E_t = \bra{\fii_t,z_t}H\ket{\fii_t,z_t}$ is the average energy and $H_{MF,S}=\bra{z}H\ket{z}$ is the mean field Hamiltonian for the system
and $H_{MF,C} = \bra{\fii}H\ket{\fii}$ is the mean field Hamiltonian for the cavity.
Clearly, state evolution preserves normalization.  By plugging in the coherent state we can verify that the coherent state indeed is a solution to the 
mean field equations. In particular, we find
\begin{align}
    \dot z &= -i\omega_c z -\frac{\kappa}{2}z-i g\langle L\rangle-if_t,\\
    \dot\phi_t &= -\omega_c\abs{z_t}^2+\frac{1}{2}g\left(\langle L^\dagger\rangle_t z_t+\langle L \rangle_t z^*_t\right)
    +\frac{1}{2}(f^*_tz_t+f_tz_t^*).
\end{align}
If the cavity damping $\kappa,\abs{f_t}\gg g$ we may adiabatically eliminate the cavity mode by setting $\dot z_t = 0$
and approximating
\begin{align}
    z_t &\approx \frac{f_t}{-\omega_c+i\frac{\kappa}{2}}.
\end{align}
This equation states that for strong classical driving and strong cavity damping, the state of the cavity remains coherent and the coherent state 
label follows the amplitude of the driving field.
The mean field evolution for the system is 
\begin{align}
   \partial_t\ket{\fii_t}=& -i (H_S+gLz^*_t +gL^\dagger z_t)\ket{\fii_t} \notag\\
   &+i(\langle H_S\rangle_t+g\langle L\rangle_t z_t^*+g\langle L^\dagger \rangle z_t)\ket{\fii_t},
\end{align}
which is non-linear. The non-linearity emerges as a global phase factor. We can define a new state $\ket{\psi_t}=e^{-iu_t}\ket{\fii_t}$
which satisfies a linear Schrödinger equation 
\begin{align}
    \partial_t\ket{\psi_t}=& -i (H_S+gLz^*_t +gL^\dagger z_t)\ket{\psi_t},
\end{align}
when $\dot u_t = \langle H_S\rangle_t+g\langle L\rangle_t z_t^*+g\langle L^\dagger \rangle z_t$.

\begin{figure*}
\begin{center}
\includegraphics[scale=0.65]{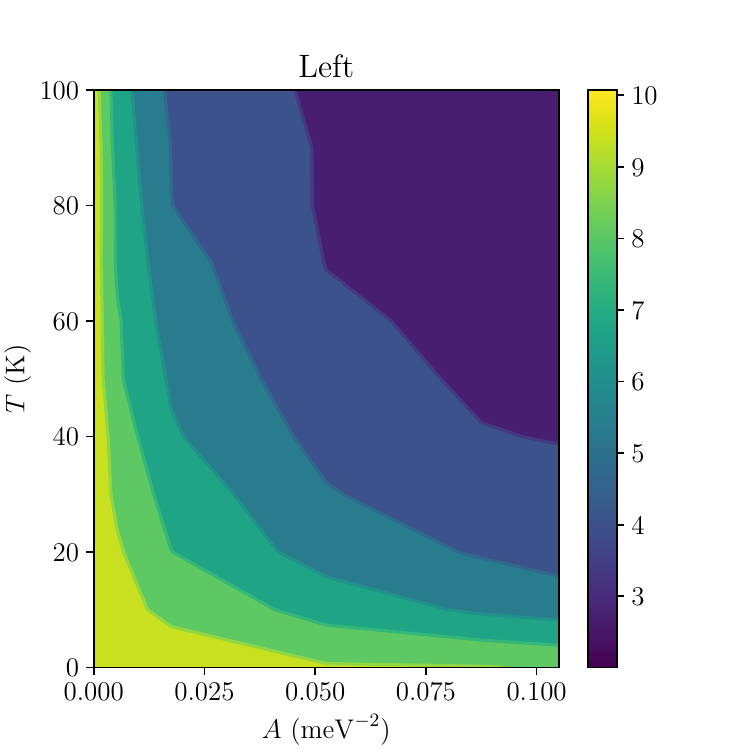}\includegraphics[scale=0.65]{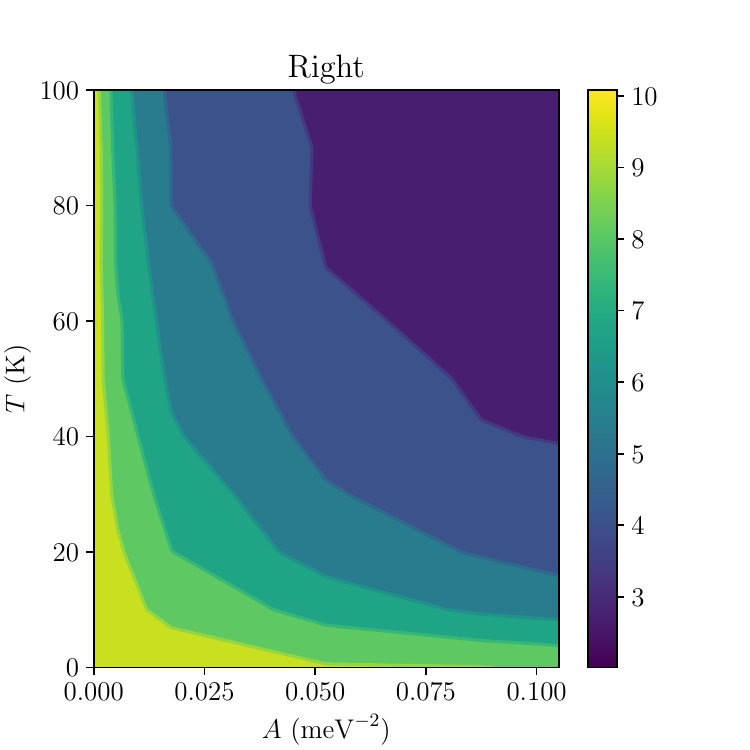}
\caption{Resolutions of the resonance fluorescence spectra side peaks as functions of temperature and coupling strength.}\label{RF_spectra_resolutions}
\end{center}
\end{figure*}

\section{Spectrum fitting}
For each of the resonance fluorescence spectra, the correlation function $\langle \tilde{\sigma}_+ \tilde{\sigma}_- (\tau) \rangle_{ss}$ was calculated using at least 60000 trajectories. At low phonon coupling strengths and low temperatures, the spectra were very smooth. However, at stronger coupling and higher temperatures, noticeable noise remained in the spectra. For the figures in the main text and the visibility analysis, we fitted three Lorentzian peaks to the numerically obtained spectra. Increasing the number of trajectories would improve the quality of the spectra from the raw data, and consequently, the fits. Figure \ref{correlation_function_raw_data_and_fit} shows the correlation function from the numerical data, the corresponding spectrum, and the fit.

\section{Spectral resolution of resonance fluorescence spectra}
The spectral resolution is defined as
\begin{align}
\mathcal{R} = \frac{\delta}{\Delta\omega},
\end{align}
where $\delta$ is the distance between the central peak and the sideband peak, and $\Delta \omega$ is the full width at half maximum (FWHM) of the central peak. Both visibility and spectral resolution quantify how well the spectral peaks can be distinguished. Higher values of either indicate that the side peaks in the resonance fluorescence spectra can be more easily identified.

The spectral resolutions for the same spectra shown in figure \ref{RF_spectra_visibilities} are presented in figure \ref{RF_spectra_resolutions}. Similar to the visibilities, the contours follow the analytic curves $T \propto 1/A$. However, a notable difference is that the spectral resolutions vary more rapidly at weak coupling strengths and low temperatures, whereas they exhibit slower variation at high temperatures and strong couplings. 
\newpage

\bibliography{references, custom_refs} % Produces the bibliography via BibTeX.

\begin{thebibliography}{53}%
\makeatletter
\providecommand \@ifxundefined [1]{%
 \@ifx{#1\undefined}
}%
\providecommand \@ifnum [1]{%
 \ifnum #1\expandafter \@firstoftwo
 \else \expandafter \@secondoftwo
 \fi
}%
\providecommand \@ifx [1]{%
 \ifx #1\expandafter \@firstoftwo
 \else \expandafter \@secondoftwo
 \fi
}%
\providecommand \natexlab [1]{#1}%
\providecommand \enquote  [1]{``#1''}%
\providecommand \bibnamefont  [1]{#1}%
\providecommand \bibfnamefont [1]{#1}%
\providecommand \citenamefont [1]{#1}%
\providecommand \href@noop [0]{\@secondoftwo}%
\providecommand \href [0]{\begingroup \@sanitize@url \@href}%
\providecommand \@href[1]{\@@startlink{#1}\@@href}%
\providecommand \@@href[1]{\endgroup#1\@@endlink}%
\providecommand \@sanitize@url [0]{\catcode `\\12\catcode `\$12\catcode `\&12\catcode `\#12\catcode `\^12\catcode `\_12\catcode `\%12\relax}%
\providecommand \@@startlink[1]{}%
\providecommand \@@endlink[0]{}%
\providecommand \url  [0]{\begingroup\@sanitize@url \@url }%
\providecommand \@url [1]{\endgroup\@href {#1}{\urlprefix }}%
\providecommand \urlprefix  [0]{URL }%
\providecommand \Eprint [0]{\href }%
\providecommand \doibase [0]{https://doi.org/}%
\providecommand \selectlanguage [0]{\@gobble}%
\providecommand \bibinfo  [0]{\@secondoftwo}%
\providecommand \bibfield  [0]{\@secondoftwo}%
\providecommand \translation [1]{[#1]}%
\providecommand \BibitemOpen [0]{}%
\providecommand \bibitemStop [0]{}%
\providecommand \bibitemNoStop [0]{.\EOS\space}%
\providecommand \EOS [0]{\spacefactor3000\relax}%
\providecommand \BibitemShut  [1]{\csname bibitem#1\endcsname}%
\let\auto@bib@innerbib\@empty
%</preamble>
\bibitem [{\citenamefont {Jordan}\ \emph {et~al.}(2024)\citenamefont {Jordan}, \citenamefont {Androvitsaneas}, \citenamefont {Clark}, \citenamefont {Trapalis}, \citenamefont {Farrer}, \citenamefont {Langbein},\ and\ \citenamefont {Bennett}}]{Jordan2024}%
  \BibitemOpen
  \bibfield  {author} {\bibinfo {author} {\bibfnamefont {M.}~\bibnamefont {Jordan}}, \bibinfo {author} {\bibfnamefont {P.}~\bibnamefont {Androvitsaneas}}, \bibinfo {author} {\bibfnamefont {R.~N.}\ \bibnamefont {Clark}}, \bibinfo {author} {\bibfnamefont {A.}~\bibnamefont {Trapalis}}, \bibinfo {author} {\bibfnamefont {I.}~\bibnamefont {Farrer}}, \bibinfo {author} {\bibfnamefont {W.}~\bibnamefont {Langbein}},\ and\ \bibinfo {author} {\bibfnamefont {A.~J.}\ \bibnamefont {Bennett}},\ }\bibfield  {title} {\bibinfo {title} {Probing purcell enhancement and photon collection efficiency of inas quantum dots at nodes of the cavity electric field},\ }\href {https://doi.org/10.1103/PhysRevResearch.6.L022004} {\bibfield  {journal} {\bibinfo  {journal} {Physical Review Research}\ }\textbf {\bibinfo {volume} {6}},\ \bibinfo {pages} {L022004} (\bibinfo {year} {2024})}\BibitemShut {NoStop}%
\bibitem [{\citenamefont {Portalupi}\ \emph {et~al.}(2015)\citenamefont {Portalupi}, \citenamefont {Hornecker}, \citenamefont {Giesz}, \citenamefont {Grange}, \citenamefont {Lemaître}, \citenamefont {Demory}, \citenamefont {Sagnes}, \citenamefont {Lanzillotti-Kimura}, \citenamefont {Lanco}, \citenamefont {Auffèves},\ and\ \citenamefont {Senellart}}]{Portalupi2015}%
  \BibitemOpen
  \bibfield  {author} {\bibinfo {author} {\bibfnamefont {S.~L.}\ \bibnamefont {Portalupi}}, \bibinfo {author} {\bibfnamefont {G.}~\bibnamefont {Hornecker}}, \bibinfo {author} {\bibfnamefont {V.}~\bibnamefont {Giesz}}, \bibinfo {author} {\bibfnamefont {T.}~\bibnamefont {Grange}}, \bibinfo {author} {\bibfnamefont {A.}~\bibnamefont {Lemaître}}, \bibinfo {author} {\bibfnamefont {J.}~\bibnamefont {Demory}}, \bibinfo {author} {\bibfnamefont {I.}~\bibnamefont {Sagnes}}, \bibinfo {author} {\bibfnamefont {N.~D.}\ \bibnamefont {Lanzillotti-Kimura}}, \bibinfo {author} {\bibfnamefont {L.}~\bibnamefont {Lanco}}, \bibinfo {author} {\bibfnamefont {A.}~\bibnamefont {Auffèves}},\ and\ \bibinfo {author} {\bibfnamefont {P.}~\bibnamefont {Senellart}},\ }\bibfield  {title} {\bibinfo {title} {Bright phonon-tuned single-photon source},\ }\href {https://doi.org/10.1021/acs.nanolett.5b00876} {\bibfield  {journal} {\bibinfo  {journal} {Nano Letters}\ }\textbf {\bibinfo {volume} {15}},\ \bibinfo {pages} {6290} (\bibinfo {year}
  {2015})}\BibitemShut {NoStop}%
\bibitem [{\citenamefont {Phillips}\ \emph {et~al.}(2024)\citenamefont {Phillips}, \citenamefont {Brash}, \citenamefont {Godsland}, \citenamefont {Martin}, \citenamefont {Foster}, \citenamefont {Tomlinson}, \citenamefont {Dost}, \citenamefont {Babazadeh}, \citenamefont {Sala}, \citenamefont {Wilson}, \citenamefont {Heffernan}, \citenamefont {Skolnick},\ and\ \citenamefont {Fox}}]{Phillips2024}%
  \BibitemOpen
  \bibfield  {author} {\bibinfo {author} {\bibfnamefont {C.~L.}\ \bibnamefont {Phillips}}, \bibinfo {author} {\bibfnamefont {A.~J.}\ \bibnamefont {Brash}}, \bibinfo {author} {\bibfnamefont {M.}~\bibnamefont {Godsland}}, \bibinfo {author} {\bibfnamefont {N.~J.}\ \bibnamefont {Martin}}, \bibinfo {author} {\bibfnamefont {A.}~\bibnamefont {Foster}}, \bibinfo {author} {\bibfnamefont {A.}~\bibnamefont {Tomlinson}}, \bibinfo {author} {\bibfnamefont {R.}~\bibnamefont {Dost}}, \bibinfo {author} {\bibfnamefont {N.}~\bibnamefont {Babazadeh}}, \bibinfo {author} {\bibfnamefont {E.~M.}\ \bibnamefont {Sala}}, \bibinfo {author} {\bibfnamefont {L.}~\bibnamefont {Wilson}}, \bibinfo {author} {\bibfnamefont {J.}~\bibnamefont {Heffernan}}, \bibinfo {author} {\bibfnamefont {M.~S.}\ \bibnamefont {Skolnick}},\ and\ \bibinfo {author} {\bibfnamefont {A.~M.}\ \bibnamefont {Fox}},\ }\bibfield  {title} {\bibinfo {title} {Purcell-enhanced single photons at telecom wavelengths from a quantum dot in a photonic crystal cavity},\ }\href
  {https://doi.org/10.1038/s41598-024-55024-6} {\bibfield  {journal} {\bibinfo  {journal} {Scientific Reports}\ }\textbf {\bibinfo {volume} {14}},\ \bibinfo {pages} {4450} (\bibinfo {year} {2024})}\BibitemShut {NoStop}%
\bibitem [{\citenamefont {Siampour}\ \emph {et~al.}(2023)\citenamefont {Siampour}, \citenamefont {O’Rourke}, \citenamefont {Brash}, \citenamefont {Makhonin}, \citenamefont {Dost}, \citenamefont {Hallett}, \citenamefont {Clarke}, \citenamefont {Patil}, \citenamefont {Skolnick},\ and\ \citenamefont {Fox}}]{Siampour2023}%
  \BibitemOpen
  \bibfield  {author} {\bibinfo {author} {\bibfnamefont {H.}~\bibnamefont {Siampour}}, \bibinfo {author} {\bibfnamefont {C.}~\bibnamefont {O’Rourke}}, \bibinfo {author} {\bibfnamefont {A.~J.}\ \bibnamefont {Brash}}, \bibinfo {author} {\bibfnamefont {M.~N.}\ \bibnamefont {Makhonin}}, \bibinfo {author} {\bibfnamefont {R.}~\bibnamefont {Dost}}, \bibinfo {author} {\bibfnamefont {D.~J.}\ \bibnamefont {Hallett}}, \bibinfo {author} {\bibfnamefont {E.}~\bibnamefont {Clarke}}, \bibinfo {author} {\bibfnamefont {P.~K.}\ \bibnamefont {Patil}}, \bibinfo {author} {\bibfnamefont {M.~S.}\ \bibnamefont {Skolnick}},\ and\ \bibinfo {author} {\bibfnamefont {A.~M.}\ \bibnamefont {Fox}},\ }\bibfield  {title} {\bibinfo {title} {Observation of large spontaneous emission rate enhancement of quantum dots in a broken-symmetry slow-light waveguide},\ }\href {https://doi.org/10.1038/s41534-023-00686-9} {\bibfield  {journal} {\bibinfo  {journal} {npj Quantum Information}\ }\textbf {\bibinfo {volume} {9}},\ \bibinfo {pages} {15}
  (\bibinfo {year} {2023})}\BibitemShut {NoStop}%
\bibitem [{\citenamefont {Daveau}\ \emph {et~al.}(2017)\citenamefont {Daveau}, \citenamefont {Balram}, \citenamefont {Pregnolato}, \citenamefont {Liu}, \citenamefont {Lee}, \citenamefont {Song}, \citenamefont {Verma}, \citenamefont {Mirin}, \citenamefont {Nam}, \citenamefont {Midolo}, \citenamefont {Stobbe}, \citenamefont {Srinivasan},\ and\ \citenamefont {Lodahl}}]{Daveau2017}%
  \BibitemOpen
  \bibfield  {author} {\bibinfo {author} {\bibfnamefont {R.~S.}\ \bibnamefont {Daveau}}, \bibinfo {author} {\bibfnamefont {K.~C.}\ \bibnamefont {Balram}}, \bibinfo {author} {\bibfnamefont {T.}~\bibnamefont {Pregnolato}}, \bibinfo {author} {\bibfnamefont {J.}~\bibnamefont {Liu}}, \bibinfo {author} {\bibfnamefont {E.~H.}\ \bibnamefont {Lee}}, \bibinfo {author} {\bibfnamefont {J.~D.}\ \bibnamefont {Song}}, \bibinfo {author} {\bibfnamefont {V.}~\bibnamefont {Verma}}, \bibinfo {author} {\bibfnamefont {R.}~\bibnamefont {Mirin}}, \bibinfo {author} {\bibfnamefont {S.~W.}\ \bibnamefont {Nam}}, \bibinfo {author} {\bibfnamefont {L.}~\bibnamefont {Midolo}}, \bibinfo {author} {\bibfnamefont {S.}~\bibnamefont {Stobbe}}, \bibinfo {author} {\bibfnamefont {K.}~\bibnamefont {Srinivasan}},\ and\ \bibinfo {author} {\bibfnamefont {P.}~\bibnamefont {Lodahl}},\ }\bibfield  {title} {\bibinfo {title} {Efficient fiber-coupled single-photon source based on quantum dots in a photonic-crystal waveguide},\ }\href
  {https://doi.org/10.1364/OPTICA.4.000178} {\bibfield  {journal} {\bibinfo  {journal} {Optica}\ }\textbf {\bibinfo {volume} {4}},\ \bibinfo {pages} {178} (\bibinfo {year} {2017})}\BibitemShut {NoStop}%
\bibitem [{\citenamefont {Giesz}\ \emph {et~al.}(2013)\citenamefont {Giesz}, \citenamefont {Gazzano}, \citenamefont {Nowak}, \citenamefont {Portalupi}, \citenamefont {Lemaître}, \citenamefont {Sagnes}, \citenamefont {Lanco},\ and\ \citenamefont {Senellart}}]{Giesz2013}%
  \BibitemOpen
  \bibfield  {author} {\bibinfo {author} {\bibfnamefont {V.}~\bibnamefont {Giesz}}, \bibinfo {author} {\bibfnamefont {O.}~\bibnamefont {Gazzano}}, \bibinfo {author} {\bibfnamefont {A.~K.}\ \bibnamefont {Nowak}}, \bibinfo {author} {\bibfnamefont {S.~L.}\ \bibnamefont {Portalupi}}, \bibinfo {author} {\bibfnamefont {A.}~\bibnamefont {Lemaître}}, \bibinfo {author} {\bibfnamefont {I.}~\bibnamefont {Sagnes}}, \bibinfo {author} {\bibfnamefont {L.}~\bibnamefont {Lanco}},\ and\ \bibinfo {author} {\bibfnamefont {P.}~\bibnamefont {Senellart}},\ }\bibfield  {title} {\bibinfo {title} {Influence of the purcell effect on the purity of bright single photon sources},\ }\bibfield  {journal} {\bibinfo  {journal} {Applied Physics Letters}\ }\textbf {\bibinfo {volume} {103}},\ \href {https://doi.org/10.1063/1.4813902} {10.1063/1.4813902} (\bibinfo {year} {2013})\BibitemShut {NoStop}%
\bibitem [{\citenamefont {Abdullah}\ \emph {et~al.}(2019)\citenamefont {Abdullah}, \citenamefont {Tang}, \citenamefont {Manolescu},\ and\ \citenamefont {Gudmundsson}}]{Abdullah2019}%
  \BibitemOpen
  \bibfield  {author} {\bibinfo {author} {\bibfnamefont {N.~R.}\ \bibnamefont {Abdullah}}, \bibinfo {author} {\bibfnamefont {C.-S.}\ \bibnamefont {Tang}}, \bibinfo {author} {\bibfnamefont {A.}~\bibnamefont {Manolescu}},\ and\ \bibinfo {author} {\bibfnamefont {V.}~\bibnamefont {Gudmundsson}},\ }\bibfield  {title} {\bibinfo {title} {Manifestation of the purcell effect in current transport through a dot–cavity–qed system},\ }\href {https://doi.org/10.3390/nano9071023} {\bibfield  {journal} {\bibinfo  {journal} {Nanomaterials}\ }\textbf {\bibinfo {volume} {9}},\ \bibinfo {pages} {1023} (\bibinfo {year} {2019})}\BibitemShut {NoStop}%
\bibitem [{\citenamefont {Liu}\ \emph {et~al.}(2018)\citenamefont {Liu}, \citenamefont {Brash}, \citenamefont {O’Hara}, \citenamefont {Martins}, \citenamefont {Phillips}, \citenamefont {Coles}, \citenamefont {Royall}, \citenamefont {Clarke}, \citenamefont {Bentham}, \citenamefont {Prtljaga}, \citenamefont {Itskevich}, \citenamefont {Wilson}, \citenamefont {Skolnick},\ and\ \citenamefont {Fox}}]{Liu2018}%
  \BibitemOpen
  \bibfield  {author} {\bibinfo {author} {\bibfnamefont {F.}~\bibnamefont {Liu}}, \bibinfo {author} {\bibfnamefont {A.~J.}\ \bibnamefont {Brash}}, \bibinfo {author} {\bibfnamefont {J.}~\bibnamefont {O’Hara}}, \bibinfo {author} {\bibfnamefont {L.~M. P.~P.}\ \bibnamefont {Martins}}, \bibinfo {author} {\bibfnamefont {C.~L.}\ \bibnamefont {Phillips}}, \bibinfo {author} {\bibfnamefont {R.~J.}\ \bibnamefont {Coles}}, \bibinfo {author} {\bibfnamefont {B.}~\bibnamefont {Royall}}, \bibinfo {author} {\bibfnamefont {E.}~\bibnamefont {Clarke}}, \bibinfo {author} {\bibfnamefont {C.}~\bibnamefont {Bentham}}, \bibinfo {author} {\bibfnamefont {N.}~\bibnamefont {Prtljaga}}, \bibinfo {author} {\bibfnamefont {I.~E.}\ \bibnamefont {Itskevich}}, \bibinfo {author} {\bibfnamefont {L.~R.}\ \bibnamefont {Wilson}}, \bibinfo {author} {\bibfnamefont {M.~S.}\ \bibnamefont {Skolnick}},\ and\ \bibinfo {author} {\bibfnamefont {A.~M.}\ \bibnamefont {Fox}},\ }\bibfield  {title} {\bibinfo {title} {High purcell factor generation of
  indistinguishable on-chip single photons},\ }\href {https://doi.org/10.1038/s41565-018-0188-x} {\bibfield  {journal} {\bibinfo  {journal} {Nature Nanotechnology}\ }\textbf {\bibinfo {volume} {13}},\ \bibinfo {pages} {835} (\bibinfo {year} {2018})}\BibitemShut {NoStop}%
\bibitem [{\citenamefont {Denning}\ \emph {et~al.}(2020)\citenamefont {Denning}, \citenamefont {Iles-Smith}, \citenamefont {Gregersen},\ and\ \citenamefont {Mork}}]{Denning2020_1}%
  \BibitemOpen
  \bibfield  {author} {\bibinfo {author} {\bibfnamefont {E.~V.}\ \bibnamefont {Denning}}, \bibinfo {author} {\bibfnamefont {J.}~\bibnamefont {Iles-Smith}}, \bibinfo {author} {\bibfnamefont {N.}~\bibnamefont {Gregersen}},\ and\ \bibinfo {author} {\bibfnamefont {J.}~\bibnamefont {Mork}},\ }\bibfield  {title} {\bibinfo {title} {Phonon effects in quantum dot single-photon sources},\ }\href {https://doi.org/10.1364/OME.380601} {\bibfield  {journal} {\bibinfo  {journal} {Optical Materials Express}\ }\textbf {\bibinfo {volume} {10}},\ \bibinfo {pages} {222} (\bibinfo {year} {2020})}\BibitemShut {NoStop}%
\bibitem [{\citenamefont {Senellart}\ \emph {et~al.}(2017)\citenamefont {Senellart}, \citenamefont {Solomon},\ and\ \citenamefont {White}}]{Senellart2017}%
  \BibitemOpen
  \bibfield  {author} {\bibinfo {author} {\bibfnamefont {P.}~\bibnamefont {Senellart}}, \bibinfo {author} {\bibfnamefont {G.}~\bibnamefont {Solomon}},\ and\ \bibinfo {author} {\bibfnamefont {A.}~\bibnamefont {White}},\ }\bibfield  {title} {\bibinfo {title} {High-performance semiconductor quantum-dot single-photon sources},\ }\href {https://doi.org/10.1038/nnano.2017.218} {\bibfield  {journal} {\bibinfo  {journal} {Nature Nanotechnology}\ }\textbf {\bibinfo {volume} {12}},\ \bibinfo {pages} {1026} (\bibinfo {year} {2017})}\BibitemShut {NoStop}%
\bibitem [{\citenamefont {Dreeßen}\ \emph {et~al.}(2018)\citenamefont {Dreeßen}, \citenamefont {Ouellet-Plamondon}, \citenamefont {Tighineanu}, \citenamefont {Zhou}, \citenamefont {Midolo}, \citenamefont {Sørensen},\ and\ \citenamefont {Lodahl}}]{Dreesen2018}%
  \BibitemOpen
  \bibfield  {author} {\bibinfo {author} {\bibfnamefont {C.~L.}\ \bibnamefont {Dreeßen}}, \bibinfo {author} {\bibfnamefont {C.}~\bibnamefont {Ouellet-Plamondon}}, \bibinfo {author} {\bibfnamefont {P.}~\bibnamefont {Tighineanu}}, \bibinfo {author} {\bibfnamefont {X.}~\bibnamefont {Zhou}}, \bibinfo {author} {\bibfnamefont {L.}~\bibnamefont {Midolo}}, \bibinfo {author} {\bibfnamefont {A.~S.}\ \bibnamefont {Sørensen}},\ and\ \bibinfo {author} {\bibfnamefont {P.}~\bibnamefont {Lodahl}},\ }\bibfield  {title} {\bibinfo {title} {Suppressing phonon decoherence of high performance single-photon sources in nanophotonic waveguides},\ }\href {https://doi.org/10.1088/2058-9565/aadbb8} {\bibfield  {journal} {\bibinfo  {journal} {Quantum Science and Technology}\ }\textbf {\bibinfo {volume} {4}},\ \bibinfo {pages} {015003} (\bibinfo {year} {2018})}\BibitemShut {NoStop}%
\bibitem [{\citenamefont {Bozzio}\ \emph {et~al.}(2022)\citenamefont {Bozzio}, \citenamefont {Vyvlecka}, \citenamefont {Cosacchi}, \citenamefont {Nawrath}, \citenamefont {Seidelmann}, \citenamefont {Loredo}, \citenamefont {Portalupi}, \citenamefont {Axt}, \citenamefont {Michler},\ and\ \citenamefont {Walther}}]{Bozzio2022}%
  \BibitemOpen
  \bibfield  {author} {\bibinfo {author} {\bibfnamefont {M.}~\bibnamefont {Bozzio}}, \bibinfo {author} {\bibfnamefont {M.}~\bibnamefont {Vyvlecka}}, \bibinfo {author} {\bibfnamefont {M.}~\bibnamefont {Cosacchi}}, \bibinfo {author} {\bibfnamefont {C.}~\bibnamefont {Nawrath}}, \bibinfo {author} {\bibfnamefont {T.}~\bibnamefont {Seidelmann}}, \bibinfo {author} {\bibfnamefont {J.~C.}\ \bibnamefont {Loredo}}, \bibinfo {author} {\bibfnamefont {S.~L.}\ \bibnamefont {Portalupi}}, \bibinfo {author} {\bibfnamefont {V.~M.}\ \bibnamefont {Axt}}, \bibinfo {author} {\bibfnamefont {P.}~\bibnamefont {Michler}},\ and\ \bibinfo {author} {\bibfnamefont {P.}~\bibnamefont {Walther}},\ }\bibfield  {title} {\bibinfo {title} {Enhancing quantum cryptography with quantum dot single-photon sources},\ }\href {https://doi.org/10.1038/s41534-022-00626-z} {\bibfield  {journal} {\bibinfo  {journal} {npj Quantum Information}\ }\textbf {\bibinfo {volume} {8}},\ \bibinfo {pages} {104} (\bibinfo {year} {2022})}\BibitemShut {NoStop}%
\bibitem [{\citenamefont {Buckley}\ \emph {et~al.}(2012)\citenamefont {Buckley}, \citenamefont {Rivoire},\ and\ \citenamefont {Vučković}}]{Buckley2012}%
  \BibitemOpen
  \bibfield  {author} {\bibinfo {author} {\bibfnamefont {S.}~\bibnamefont {Buckley}}, \bibinfo {author} {\bibfnamefont {K.}~\bibnamefont {Rivoire}},\ and\ \bibinfo {author} {\bibfnamefont {J.}~\bibnamefont {Vučković}},\ }\bibfield  {title} {\bibinfo {title} {Engineered quantum dot single-photon sources},\ }\href {https://doi.org/10.1088/0034-4885/75/12/126503} {\bibfield  {journal} {\bibinfo  {journal} {Reports on Progress in Physics}\ }\textbf {\bibinfo {volume} {75}},\ \bibinfo {pages} {126503} (\bibinfo {year} {2012})}\BibitemShut {NoStop}%
\bibitem [{\citenamefont {Gustin}\ and\ \citenamefont {Hughes}(2018)}]{Gustin2018}%
  \BibitemOpen
  \bibfield  {author} {\bibinfo {author} {\bibfnamefont {C.}~\bibnamefont {Gustin}}\ and\ \bibinfo {author} {\bibfnamefont {S.}~\bibnamefont {Hughes}},\ }\bibfield  {title} {\bibinfo {title} {Pulsed excitation dynamics in quantum-dot–cavity systems: Limits to optimizing the fidelity of on-demand single-photon sources},\ }\href {https://doi.org/10.1103/PhysRevB.98.045309} {\bibfield  {journal} {\bibinfo  {journal} {Physical Review B}\ }\textbf {\bibinfo {volume} {98}},\ \bibinfo {pages} {045309} (\bibinfo {year} {2018})}\BibitemShut {NoStop}%
\bibitem [{\citenamefont {Lindblad}(1976)}]{Lindblad1976}%
  \BibitemOpen
  \bibfield  {author} {\bibinfo {author} {\bibfnamefont {G.}~\bibnamefont {Lindblad}},\ }\bibfield  {title} {\bibinfo {title} {On the generators of quantum dynamical semigroups},\ }\href {https://doi.org/10.1007/BF01608499} {\bibfield  {journal} {\bibinfo  {journal} {Communications in Mathematical Physics}\ }\textbf {\bibinfo {volume} {48}},\ \bibinfo {pages} {119} (\bibinfo {year} {1976})}\BibitemShut {NoStop}%
\bibitem [{\citenamefont {Gorini}\ \emph {et~al.}(1976)\citenamefont {Gorini}, \citenamefont {Kossakowski},\ and\ \citenamefont {Sudarshan}}]{Gorini1976}%
  \BibitemOpen
  \bibfield  {author} {\bibinfo {author} {\bibfnamefont {V.}~\bibnamefont {Gorini}}, \bibinfo {author} {\bibfnamefont {A.}~\bibnamefont {Kossakowski}},\ and\ \bibinfo {author} {\bibfnamefont {E.~C.~G.}\ \bibnamefont {Sudarshan}},\ }\bibfield  {title} {\bibinfo {title} {Completely positive dynamical semigroups of <i>n</i> -level systems},\ }\href {https://doi.org/10.1063/1.522979} {\bibfield  {journal} {\bibinfo  {journal} {Journal of Mathematical Physics}\ }\textbf {\bibinfo {volume} {17}},\ \bibinfo {pages} {821} (\bibinfo {year} {1976})}\BibitemShut {NoStop}%
\bibitem [{\citenamefont {Carmichael}(1993)}]{Carmichael1993}%
  \BibitemOpen
  \bibfield  {author} {\bibinfo {author} {\bibfnamefont {H.}~\bibnamefont {Carmichael}},\ }\href {https://doi.org/10.1007/978-3-540-47620-7} {\emph {\bibinfo {title} {An Open Systems Approach to Quantum Optics}}},\ Vol.~\bibinfo {volume} {18}\ (\bibinfo  {publisher} {Springer Berlin Heidelberg},\ \bibinfo {year} {1993})\BibitemShut {NoStop}%
\bibitem [{\citenamefont {Gardiner}\ and\ \citenamefont {Zoller}(2004)}]{Gardiner2004}%
  \BibitemOpen
  \bibfield  {author} {\bibinfo {author} {\bibfnamefont {C.~W.}\ \bibnamefont {Gardiner}}\ and\ \bibinfo {author} {\bibfnamefont {P.}~\bibnamefont {Zoller}},\ }\href@noop {} {\emph {\bibinfo {title} {Quantum noise: a handbook of Markovian and non-Markovian quantum stochastic methods with applications to quantum optics}}},\ \bibinfo {edition} {2nd}\ ed.,\ Vol.~\bibinfo {volume} {16}\ (\bibinfo  {publisher} {Springer},\ \bibinfo {year} {2004})\BibitemShut {NoStop}%
\bibitem [{\citenamefont {Breuer}\ and\ \citenamefont {Petruccione}(2007)}]{Breuer2007}%
  \BibitemOpen
  \bibfield  {author} {\bibinfo {author} {\bibfnamefont {H.-P.}\ \bibnamefont {Breuer}}\ and\ \bibinfo {author} {\bibfnamefont {F.}~\bibnamefont {Petruccione}},\ }\href {https://doi.org/10.1093/acprof:oso/9780199213900.001.0001} {\emph {\bibinfo {title} {The Theory of Open Quantum Systems}}}\ (\bibinfo  {publisher} {Oxford University PressOxford},\ \bibinfo {year} {2007})\BibitemShut {NoStop}%
\bibitem [{\citenamefont {Guarnieri}\ \emph {et~al.}(2014)\citenamefont {Guarnieri}, \citenamefont {Smirne},\ and\ \citenamefont {Vacchini}}]{Guarnieri2014}%
  \BibitemOpen
  \bibfield  {author} {\bibinfo {author} {\bibfnamefont {G.}~\bibnamefont {Guarnieri}}, \bibinfo {author} {\bibfnamefont {A.}~\bibnamefont {Smirne}},\ and\ \bibinfo {author} {\bibfnamefont {B.}~\bibnamefont {Vacchini}},\ }\bibfield  {title} {\bibinfo {title} {Quantum regression theorem and non-markovianity of quantum dynamics},\ }\href {https://doi.org/10.1103/PhysRevA.90.022110} {\bibfield  {journal} {\bibinfo  {journal} {Physical Review A}\ }\textbf {\bibinfo {volume} {90}},\ \bibinfo {pages} {022110} (\bibinfo {year} {2014})}\BibitemShut {NoStop}%
\bibitem [{\citenamefont {Cosacchi}\ \emph {et~al.}(2021)\citenamefont {Cosacchi}, \citenamefont {Seidelmann}, \citenamefont {Cygorek}, \citenamefont {Vagov}, \citenamefont {Reiter},\ and\ \citenamefont {Axt}}]{Cosacchi2021}%
  \BibitemOpen
  \bibfield  {author} {\bibinfo {author} {\bibfnamefont {M.}~\bibnamefont {Cosacchi}}, \bibinfo {author} {\bibfnamefont {T.}~\bibnamefont {Seidelmann}}, \bibinfo {author} {\bibfnamefont {M.}~\bibnamefont {Cygorek}}, \bibinfo {author} {\bibfnamefont {A.}~\bibnamefont {Vagov}}, \bibinfo {author} {\bibfnamefont {D.}~\bibnamefont {Reiter}},\ and\ \bibinfo {author} {\bibfnamefont {V.}~\bibnamefont {Axt}},\ }\bibfield  {title} {\bibinfo {title} {Accuracy of the quantum regression theorem for photon emission from a quantum dot},\ }\href {https://doi.org/10.1103/PhysRevLett.127.100402} {\bibfield  {journal} {\bibinfo  {journal} {Physical Review Letters}\ }\textbf {\bibinfo {volume} {127}},\ \bibinfo {pages} {100402} (\bibinfo {year} {2021})}\BibitemShut {NoStop}%
\bibitem [{\citenamefont {Khan}\ \emph {et~al.}(2024)\citenamefont {Khan}, \citenamefont {Agarwalla},\ and\ \citenamefont {Jain}}]{Khan2024}%
  \BibitemOpen
  \bibfield  {author} {\bibinfo {author} {\bibfnamefont {S.}~\bibnamefont {Khan}}, \bibinfo {author} {\bibfnamefont {B.~K.}\ \bibnamefont {Agarwalla}},\ and\ \bibinfo {author} {\bibfnamefont {S.}~\bibnamefont {Jain}},\ }\bibfield  {title} {\bibinfo {title} {Modified quantum regression theorem and consistency with kubo-martin-schwinger condition},\ }\href {https://doi.org/10.1088/1367-2630/ad976f} {\bibfield  {journal} {\bibinfo  {journal} {New Journal of Physics}\ }\textbf {\bibinfo {volume} {26}},\ \bibinfo {pages} {123011} (\bibinfo {year} {2024})}\BibitemShut {NoStop}%
\bibitem [{\citenamefont {Khan}\ \emph {et~al.}(2022)\citenamefont {Khan}, \citenamefont {Agarwalla},\ and\ \citenamefont {Jain}}]{Khan2022}%
  \BibitemOpen
  \bibfield  {author} {\bibinfo {author} {\bibfnamefont {S.}~\bibnamefont {Khan}}, \bibinfo {author} {\bibfnamefont {B.~K.}\ \bibnamefont {Agarwalla}},\ and\ \bibinfo {author} {\bibfnamefont {S.}~\bibnamefont {Jain}},\ }\bibfield  {title} {\bibinfo {title} {Quantum regression theorem for multi-time correlators: A detailed analysis in the heisenberg picture},\ }\href {https://doi.org/10.1103/PhysRevA.106.022214} {\bibfield  {journal} {\bibinfo  {journal} {Physical Review A}\ }\textbf {\bibinfo {volume} {106}},\ \bibinfo {pages} {022214} (\bibinfo {year} {2022})}\BibitemShut {NoStop}%
\bibitem [{\citenamefont {Lonigro}\ and\ \citenamefont {Chruściński}(2022)}]{Lonigro2022}%
  \BibitemOpen
  \bibfield  {author} {\bibinfo {author} {\bibfnamefont {D.}~\bibnamefont {Lonigro}}\ and\ \bibinfo {author} {\bibfnamefont {D.}~\bibnamefont {Chruściński}},\ }\bibfield  {title} {\bibinfo {title} {Quantum regression beyond the born-markov approximation for generalized spin-boson models},\ }\href {https://doi.org/10.1103/PhysRevA.105.052435} {\bibfield  {journal} {\bibinfo  {journal} {Physical Review A}\ }\textbf {\bibinfo {volume} {105}},\ \bibinfo {pages} {052435} (\bibinfo {year} {2022})}\BibitemShut {NoStop}%
\bibitem [{\citenamefont {Panyukov}\ \emph {et~al.}(2024)\citenamefont {Panyukov}, \citenamefont {Shishkov},\ and\ \citenamefont {Andrianov}}]{Panyukov2024}%
  \BibitemOpen
  \bibfield  {author} {\bibinfo {author} {\bibfnamefont {I.~V.}\ \bibnamefont {Panyukov}}, \bibinfo {author} {\bibfnamefont {V.~Y.}\ \bibnamefont {Shishkov}},\ and\ \bibinfo {author} {\bibfnamefont {E.~S.}\ \bibnamefont {Andrianov}},\ }\bibfield  {title} {\bibinfo {title} {Adjoint master equation for multitime correlators},\ }\href {https://doi.org/10.1103/PhysRevA.109.052207} {\bibfield  {journal} {\bibinfo  {journal} {Physical Review A}\ }\textbf {\bibinfo {volume} {109}},\ \bibinfo {pages} {052207} (\bibinfo {year} {2024})}\BibitemShut {NoStop}%
\bibitem [{\citenamefont {Boedecker}\ and\ \citenamefont {Henkel}(2012)}]{Boedecker2012}%
  \BibitemOpen
  \bibfield  {author} {\bibinfo {author} {\bibfnamefont {G.}~\bibnamefont {Boedecker}}\ and\ \bibinfo {author} {\bibfnamefont {C.}~\bibnamefont {Henkel}},\ }\bibfield  {title} {\bibinfo {title} {Validity of the quantum regression theorem for resonance fluorescence in a photonic crystal},\ }\href {https://doi.org/10.1002/andp.201200135} {\bibfield  {journal} {\bibinfo  {journal} {Annalen der Physik}\ }\textbf {\bibinfo {volume} {524}},\ \bibinfo {pages} {805} (\bibinfo {year} {2012})}\BibitemShut {NoStop}%
\bibitem [{\citenamefont {Talkner}(1986)}]{Talkner1986}%
  \BibitemOpen
  \bibfield  {author} {\bibinfo {author} {\bibfnamefont {P.}~\bibnamefont {Talkner}},\ }\bibfield  {title} {\bibinfo {title} {The failure of the quantum regression hypothesis},\ }\href {https://doi.org/10.1016/0003-4916(86)90207-1} {\bibfield  {journal} {\bibinfo  {journal} {Annals of Physics}\ }\textbf {\bibinfo {volume} {167}},\ \bibinfo {pages} {390} (\bibinfo {year} {1986})}\BibitemShut {NoStop}%
\bibitem [{\citenamefont {Alonso}\ and\ \citenamefont {de~Vega}(2005)}]{Alonso2005}%
  \BibitemOpen
  \bibfield  {author} {\bibinfo {author} {\bibfnamefont {D.}~\bibnamefont {Alonso}}\ and\ \bibinfo {author} {\bibfnamefont {I.}~\bibnamefont {de~Vega}},\ }\bibfield  {title} {\bibinfo {title} {Multiple-time correlation functions for non-markovian interaction: Beyond the quantum regression theorem},\ }\href {https://doi.org/10.1103/PhysRevLett.94.200403} {\bibfield  {journal} {\bibinfo  {journal} {Physical Review Letters}\ }\textbf {\bibinfo {volume} {94}},\ \bibinfo {pages} {200403} (\bibinfo {year} {2005})}\BibitemShut {NoStop}%
\bibitem [{\citenamefont {Tarasov}(2021)}]{Tarasov2021}%
  \BibitemOpen
  \bibfield  {author} {\bibinfo {author} {\bibfnamefont {V.~E.}\ \bibnamefont {Tarasov}},\ }\bibfield  {title} {\bibinfo {title} {General non-markovian quantum dynamics},\ }\href {https://doi.org/10.3390/e23081006} {\bibfield  {journal} {\bibinfo  {journal} {Entropy}\ }\textbf {\bibinfo {volume} {23}},\ \bibinfo {pages} {1006} (\bibinfo {year} {2021})}\BibitemShut {NoStop}%
\bibitem [{\citenamefont {Dowling}\ \emph {et~al.}(2023)\citenamefont {Dowling}, \citenamefont {Figueroa-Romero}, \citenamefont {Pollock}, \citenamefont {Strasberg},\ and\ \citenamefont {Modi}}]{Dowling2023}%
  \BibitemOpen
  \bibfield  {author} {\bibinfo {author} {\bibfnamefont {N.}~\bibnamefont {Dowling}}, \bibinfo {author} {\bibfnamefont {P.}~\bibnamefont {Figueroa-Romero}}, \bibinfo {author} {\bibfnamefont {F.~A.}\ \bibnamefont {Pollock}}, \bibinfo {author} {\bibfnamefont {P.}~\bibnamefont {Strasberg}},\ and\ \bibinfo {author} {\bibfnamefont {K.}~\bibnamefont {Modi}},\ }\bibfield  {title} {\bibinfo {title} {Relaxation of multitime statistics in quantum systems},\ }\href {https://doi.org/10.22331/q-2023-06-01-1027} {\bibfield  {journal} {\bibinfo  {journal} {Quantum}\ }\textbf {\bibinfo {volume} {7}},\ \bibinfo {pages} {1027} (\bibinfo {year} {2023})}\BibitemShut {NoStop}%
\bibitem [{\citenamefont {Burgarth}\ \emph {et~al.}(2021)\citenamefont {Burgarth}, \citenamefont {Facchi}, \citenamefont {Lonigro},\ and\ \citenamefont {Modi}}]{Burgarth2021}%
  \BibitemOpen
  \bibfield  {author} {\bibinfo {author} {\bibfnamefont {D.}~\bibnamefont {Burgarth}}, \bibinfo {author} {\bibfnamefont {P.}~\bibnamefont {Facchi}}, \bibinfo {author} {\bibfnamefont {D.}~\bibnamefont {Lonigro}},\ and\ \bibinfo {author} {\bibfnamefont {K.}~\bibnamefont {Modi}},\ }\bibfield  {title} {\bibinfo {title} {Quantum non-markovianity elusive to interventions},\ }\href {https://doi.org/10.1103/PhysRevA.104.L050404} {\bibfield  {journal} {\bibinfo  {journal} {Physical Review A}\ }\textbf {\bibinfo {volume} {104}},\ \bibinfo {pages} {L050404} (\bibinfo {year} {2021})}\BibitemShut {NoStop}%
\bibitem [{\citenamefont {Zambon}\ and\ \citenamefont {Soares-Pinto}(2024)}]{Zambon2024}%
  \BibitemOpen
  \bibfield  {author} {\bibinfo {author} {\bibfnamefont {G.}~\bibnamefont {Zambon}}\ and\ \bibinfo {author} {\bibfnamefont {D.~O.}\ \bibnamefont {Soares-Pinto}},\ }\bibfield  {title} {\bibinfo {title} {Relations between markovian and non-markovian correlations in multitime quantum processes},\ }\href {https://doi.org/10.1103/PhysRevA.109.062401} {\bibfield  {journal} {\bibinfo  {journal} {Physical Review A}\ }\textbf {\bibinfo {volume} {109}},\ \bibinfo {pages} {062401} (\bibinfo {year} {2024})}\BibitemShut {NoStop}%
\bibitem [{\citenamefont {Smirne}\ \emph {et~al.}(2022)\citenamefont {Smirne}, \citenamefont {Tamascelli}, \citenamefont {Lim}, \citenamefont {Plenio},\ and\ \citenamefont {Huelga}}]{Smirne2022}%
  \BibitemOpen
  \bibfield  {author} {\bibinfo {author} {\bibfnamefont {A.}~\bibnamefont {Smirne}}, \bibinfo {author} {\bibfnamefont {D.}~\bibnamefont {Tamascelli}}, \bibinfo {author} {\bibfnamefont {J.}~\bibnamefont {Lim}}, \bibinfo {author} {\bibfnamefont {M.~B.}\ \bibnamefont {Plenio}},\ and\ \bibinfo {author} {\bibfnamefont {S.~F.}\ \bibnamefont {Huelga}},\ }\bibfield  {title} {\bibinfo {title} {Non-perturbative treatment of open-system multi-time expectation values in gaussian bosonic environments},\ }\bibfield  {journal} {\bibinfo  {journal} {Open Systems \& Information Dynamics}\ }\textbf {\bibinfo {volume} {29}},\ \href {https://doi.org/10.1142/S1230161222500196} {10.1142/S1230161222500196} (\bibinfo {year} {2022})\BibitemShut {NoStop}%
\bibitem [{\citenamefont {Campaioli}\ \emph {et~al.}(2024)\citenamefont {Campaioli}, \citenamefont {Cole},\ and\ \citenamefont {Hapuarachchi}}]{Campaioli2024}%
  \BibitemOpen
  \bibfield  {author} {\bibinfo {author} {\bibfnamefont {F.}~\bibnamefont {Campaioli}}, \bibinfo {author} {\bibfnamefont {J.~H.}\ \bibnamefont {Cole}},\ and\ \bibinfo {author} {\bibfnamefont {H.}~\bibnamefont {Hapuarachchi}},\ }\bibfield  {title} {\bibinfo {title} {Quantum master equations: Tips and tricks for quantum optics, quantum computing, and beyond},\ }\href {https://doi.org/10.1103/PRXQuantum.5.020202} {\bibfield  {journal} {\bibinfo  {journal} {PRX Quantum}\ }\textbf {\bibinfo {volume} {5}},\ \bibinfo {pages} {020202} (\bibinfo {year} {2024})}\BibitemShut {NoStop}%
\bibitem [{\citenamefont {Strathearn}\ \emph {et~al.}(2018)\citenamefont {Strathearn}, \citenamefont {Kirton}, \citenamefont {Kilda}, \citenamefont {Keeling},\ and\ \citenamefont {Lovett}}]{Starthearn2018}%
  \BibitemOpen
  \bibfield  {author} {\bibinfo {author} {\bibfnamefont {A.}~\bibnamefont {Strathearn}}, \bibinfo {author} {\bibfnamefont {P.}~\bibnamefont {Kirton}}, \bibinfo {author} {\bibfnamefont {D.}~\bibnamefont {Kilda}}, \bibinfo {author} {\bibfnamefont {J.}~\bibnamefont {Keeling}},\ and\ \bibinfo {author} {\bibfnamefont {B.~W.}\ \bibnamefont {Lovett}},\ }\bibfield  {title} {\bibinfo {title} {Efficient non-markovian quantum dynamics using time-evolving matrix product operators},\ }\href {https://doi.org/10.1038/s41467-018-05617-3} {\bibfield  {journal} {\bibinfo  {journal} {Nature Communications}\ }\textbf {\bibinfo {volume} {9}},\ \bibinfo {pages} {3322} (\bibinfo {year} {2018})}\BibitemShut {NoStop}%
\bibitem [{\citenamefont {Shrikant}\ and\ \citenamefont {Mandayam}(2023)}]{Shurikant2023}%
  \BibitemOpen
  \bibfield  {author} {\bibinfo {author} {\bibfnamefont {U.}~\bibnamefont {Shrikant}}\ and\ \bibinfo {author} {\bibfnamefont {P.}~\bibnamefont {Mandayam}},\ }\bibfield  {title} {\bibinfo {title} {Quantum non-markovianity: Overview and recent developments},\ }\bibfield  {journal} {\bibinfo  {journal} {Frontiers in Quantum Science and Technology}\ }\textbf {\bibinfo {volume} {2}},\ \href {https://doi.org/10.3389/frqst.2023.1134583} {10.3389/frqst.2023.1134583} (\bibinfo {year} {2023})\BibitemShut {NoStop}%
\bibitem [{\citenamefont {Ziman}(2001)}]{Ziman2001}%
  \BibitemOpen
  \bibfield  {author} {\bibinfo {author} {\bibfnamefont {J.}~\bibnamefont {Ziman}},\ }\href {https://doi.org/10.1093/acprof:oso/9780198507796.001.0001} {\emph {\bibinfo {title} {Electrons and Phonons}}}\ (\bibinfo  {publisher} {Oxford University Press},\ \bibinfo {year} {2001})\BibitemShut {NoStop}%
\bibitem [{\citenamefont {Suess}\ \emph {et~al.}(2014)\citenamefont {Suess}, \citenamefont {Eisfeld},\ and\ \citenamefont {Strunz}}]{Suess2014}%
  \BibitemOpen
  \bibfield  {author} {\bibinfo {author} {\bibfnamefont {D.}~\bibnamefont {Suess}}, \bibinfo {author} {\bibfnamefont {A.}~\bibnamefont {Eisfeld}},\ and\ \bibinfo {author} {\bibfnamefont {W.~T.}\ \bibnamefont {Strunz}},\ }\bibfield  {title} {\bibinfo {title} {Hierarchy of stochastic pure states for open quantum system dynamics},\ }\bibfield  {journal} {\bibinfo  {journal} {Physical Review Letters}\ }\textbf {\bibinfo {volume} {113}},\ \href {https://doi.org/10.1103/PhysRevLett.113.150403} {10.1103/PhysRevLett.113.150403} (\bibinfo {year} {2014})\BibitemShut {NoStop}%
\bibitem [{\citenamefont {Diosi}\ and\ \citenamefont {Strunz}(1997)}]{Diosi1997}%
  \BibitemOpen
  \bibfield  {author} {\bibinfo {author} {\bibfnamefont {L.}~\bibnamefont {Diosi}}\ and\ \bibinfo {author} {\bibfnamefont {W.~T.}\ \bibnamefont {Strunz}},\ }\bibfield  {title} {\bibinfo {title} {The non-markovian stochastic schrödinger equation for open systems},\ }\href {https://doi.org/10.1016/S0375-9601(97)00717-2} {\bibfield  {journal} {\bibinfo  {journal} {Physics Letters A}\ }\textbf {\bibinfo {volume} {235}},\ \bibinfo {pages} {569} (\bibinfo {year} {1997})}\BibitemShut {NoStop}%
\bibitem [{\citenamefont {Diósi}\ \emph {et~al.}(1998)\citenamefont {Diósi}, \citenamefont {Gisin},\ and\ \citenamefont {Strunz}}]{Diosi1998}%
  \BibitemOpen
  \bibfield  {author} {\bibinfo {author} {\bibfnamefont {L.}~\bibnamefont {Diósi}}, \bibinfo {author} {\bibfnamefont {N.}~\bibnamefont {Gisin}},\ and\ \bibinfo {author} {\bibfnamefont {W.~T.}\ \bibnamefont {Strunz}},\ }\bibfield  {title} {\bibinfo {title} {Non-markovian quantum state diffusion},\ }\href {https://doi.org/10.1103/PhysRevA.58.1699} {\bibfield  {journal} {\bibinfo  {journal} {Physical Review A}\ }\textbf {\bibinfo {volume} {58}},\ \bibinfo {pages} {1699} (\bibinfo {year} {1998})},\ \bibinfo {note} {tärkeä<br/>}\BibitemShut {NoStop}%
\bibitem [{\citenamefont {Hartmann}\ and\ \citenamefont {Strunz}(2017)}]{Hartmann2017}%
  \BibitemOpen
  \bibfield  {author} {\bibinfo {author} {\bibfnamefont {R.}~\bibnamefont {Hartmann}}\ and\ \bibinfo {author} {\bibfnamefont {W.~T.}\ \bibnamefont {Strunz}},\ }\bibfield  {title} {\bibinfo {title} {Exact open quantum system dynamics using the hierarchy of pure states (hops)},\ }\href {https://doi.org/10.1021/acs.jctc.7b00751} {\bibfield  {journal} {\bibinfo  {journal} {Journal of Chemical Theory and Computation}\ }\textbf {\bibinfo {volume} {13}},\ \bibinfo {pages} {5834} (\bibinfo {year} {2017})}\BibitemShut {NoStop}%
\bibitem [{\citenamefont {Goetsch}\ \emph {et~al.}(1996)\citenamefont {Goetsch}, \citenamefont {Graham},\ and\ \citenamefont {Haake}}]{Goetsch1996}%
  \BibitemOpen
  \bibfield  {author} {\bibinfo {author} {\bibfnamefont {P.}~\bibnamefont {Goetsch}}, \bibinfo {author} {\bibfnamefont {R.}~\bibnamefont {Graham}},\ and\ \bibinfo {author} {\bibfnamefont {F.}~\bibnamefont {Haake}},\ }\bibfield  {title} {\bibinfo {title} {Microscopic foundation of a finite-temperature stochastic schrödinger equation},\ }\href {https://doi.org/10.1088/1355-5111/8/1/012} {\bibfield  {journal} {\bibinfo  {journal} {Quantum and Semiclassical Optics: Journal of the European Optical Society Part B}\ }\textbf {\bibinfo {volume} {8}},\ \bibinfo {pages} {157} (\bibinfo {year} {1996})}\BibitemShut {NoStop}%
\bibitem [{\citenamefont {Hartmann}\ and\ \citenamefont {Strunz}(2021)}]{Hartmann2021}%
  \BibitemOpen
  \bibfield  {author} {\bibinfo {author} {\bibfnamefont {R.}~\bibnamefont {Hartmann}}\ and\ \bibinfo {author} {\bibfnamefont {W.~T.}\ \bibnamefont {Strunz}},\ }\bibfield  {title} {\bibinfo {title} {Open quantum system response from the hierarchy of pure states},\ }\href {https://doi.org/10.1021/acs.jpca.1c03339} {\bibfield  {journal} {\bibinfo  {journal} {The Journal of Physical Chemistry A}\ }\textbf {\bibinfo {volume} {125}},\ \bibinfo {pages} {7066} (\bibinfo {year} {2021})}\BibitemShut {NoStop}%
\bibitem [{\citenamefont {Kok}\ and\ \citenamefont {Lovett}(2010)}]{Kok2010}%
  \BibitemOpen
  \bibfield  {author} {\bibinfo {author} {\bibfnamefont {P.}~\bibnamefont {Kok}}\ and\ \bibinfo {author} {\bibfnamefont {B.~W.}\ \bibnamefont {Lovett}},\ }\href {https://doi.org/10.1017/CBO9781139193658} {\emph {\bibinfo {title} {Introduction to Optical Quantum Information Processing}}}\ (\bibinfo  {publisher} {Cambridge University Press},\ \bibinfo {year} {2010})\BibitemShut {NoStop}%
\bibitem [{\citenamefont {Mahan}(2000)}]{Mahan2000}%
  \BibitemOpen
  \bibfield  {author} {\bibinfo {author} {\bibfnamefont {G.~D.}\ \bibnamefont {Mahan}},\ }\href {https://doi.org/10.1007/978-1-4757-5714-9} {\emph {\bibinfo {title} {Many-Particle Physics}}}\ (\bibinfo  {publisher} {Springer US},\ \bibinfo {year} {2000})\BibitemShut {NoStop}%
\bibitem [{\citenamefont {Krummheuer}\ \emph {et~al.}(2002)\citenamefont {Krummheuer}, \citenamefont {Axt},\ and\ \citenamefont {Kuhn}}]{Krummheuer2002}%
  \BibitemOpen
  \bibfield  {author} {\bibinfo {author} {\bibfnamefont {B.}~\bibnamefont {Krummheuer}}, \bibinfo {author} {\bibfnamefont {V.~M.}\ \bibnamefont {Axt}},\ and\ \bibinfo {author} {\bibfnamefont {T.}~\bibnamefont {Kuhn}},\ }\bibfield  {title} {\bibinfo {title} {Theory of pure dephasing and the resulting absorption line shape in semiconductor quantum dots},\ }\href {https://doi.org/10.1103/PhysRevB.65.195313} {\bibfield  {journal} {\bibinfo  {journal} {Physical Review B}\ }\textbf {\bibinfo {volume} {65}},\ \bibinfo {pages} {195313} (\bibinfo {year} {2002})}\BibitemShut {NoStop}%
\bibitem [{\citenamefont {Hohenester}(2007)}]{Hohenster2007}%
  \BibitemOpen
  \bibfield  {author} {\bibinfo {author} {\bibfnamefont {U.}~\bibnamefont {Hohenester}},\ }\bibfield  {title} {\bibinfo {title} {Quantum control of polaron states in semiconductor quantum dots},\ }\href {https://doi.org/10.1088/0953-4075/40/11/S06} {\bibfield  {journal} {\bibinfo  {journal} {Journal of Physics B: Atomic, Molecular and Optical Physics}\ }\textbf {\bibinfo {volume} {40}},\ \bibinfo {pages} {S315} (\bibinfo {year} {2007})}\BibitemShut {NoStop}%
\bibitem [{\citenamefont {Toivonen}(2023)}]{Toivonen2023}%
  \BibitemOpen
  \bibfield  {author} {\bibinfo {author} {\bibfnamefont {S.}~\bibnamefont {Toivonen}},\ }\emph {\bibinfo {title} {Simulating the Dynamics of Quantum Dots using Hierarchy of Pure States}},\ \href@noop {} {\bibinfo {type} {Master's thesis}},\ \bibinfo  {school} {University of Turku}, \bibinfo {address} {Department of Physics and Astronomy} (\bibinfo {year} {2023})\BibitemShut {NoStop}%
\bibitem [{\citenamefont {May}\ and\ \citenamefont {Kühn}(2003)}]{Kuhn2003}%
  \BibitemOpen
  \bibfield  {author} {\bibinfo {author} {\bibfnamefont {V.}~\bibnamefont {May}}\ and\ \bibinfo {author} {\bibfnamefont {O.}~\bibnamefont {Kühn}},\ }\href {https://doi.org/10.1002/9783527602575} {\emph {\bibinfo {title} {Charge and Energy Transfer Dynamics in Molecular Systems}}}\ (\bibinfo  {publisher} {Wiley},\ \bibinfo {year} {2003})\BibitemShut {NoStop}%
\bibitem [{\citenamefont {Roden}\ \emph {et~al.}(2011)\citenamefont {Roden}, \citenamefont {Strunz},\ and\ \citenamefont {Eisfeld}}]{Roden2011}%
  \BibitemOpen
  \bibfield  {author} {\bibinfo {author} {\bibfnamefont {J.}~\bibnamefont {Roden}}, \bibinfo {author} {\bibfnamefont {W.~T.}\ \bibnamefont {Strunz}},\ and\ \bibinfo {author} {\bibfnamefont {A.}~\bibnamefont {Eisfeld}},\ }\bibfield  {title} {\bibinfo {title} {Non-markovian quantum state diffusion for absorption spectra of molecular aggregates},\ }\bibfield  {journal} {\bibinfo  {journal} {Journal of Chemical Physics}\ }\textbf {\bibinfo {volume} {134}},\ \href {https://doi.org/10.1063/1.3512979} {10.1063/1.3512979} (\bibinfo {year} {2011})\BibitemShut {NoStop}%
\bibitem [{\citenamefont {Roy}\ and\ \citenamefont {Hughes}(2012)}]{Roy2012}%
  \BibitemOpen
  \bibfield  {author} {\bibinfo {author} {\bibfnamefont {C.}~\bibnamefont {Roy}}\ and\ \bibinfo {author} {\bibfnamefont {S.}~\bibnamefont {Hughes}},\ }\bibfield  {title} {\bibinfo {title} {Polaron master equation theory of the quantum-dot mollow triplet in a semiconductor cavity-qed system},\ }\href {https://doi.org/10.1103/PhysRevB.85.115309} {\bibfield  {journal} {\bibinfo  {journal} {Physical Review B}\ }\textbf {\bibinfo {volume} {85}},\ \bibinfo {pages} {115309} (\bibinfo {year} {2012})}\BibitemShut {NoStop}%
\bibitem [{ste()}]{steck}%
  \BibitemOpen
  \href@noop {} {}\bibinfo {note} {Daniel A. Steck, \textit{Quantum and Atom Optics}, available online at \url{http://steck.us/teaching} (revision 0.16.2, 15 November 2024).}\BibitemShut {Stop}%
\bibitem [{\citenamefont {Kümmerer}\ and\ \citenamefont {Maassen}(2004)}]{kummerer2004}%
  \BibitemOpen
  \bibfield  {author} {\bibinfo {author} {\bibfnamefont {B.}~\bibnamefont {Kümmerer}}\ and\ \bibinfo {author} {\bibfnamefont {H.}~\bibnamefont {Maassen}},\ }\bibfield  {title} {\bibinfo {title} {A pathwise ergodic theorem for quantum trajectories},\ }\href {https://doi.org/10.1088/0305-4470/37/49/008} {\bibfield  {journal} {\bibinfo  {journal} {Journal of Physics A: Mathematical and General}\ }\textbf {\bibinfo {volume} {37}},\ \bibinfo {pages} {11889} (\bibinfo {year} {2004})}\BibitemShut {NoStop}%
\end{thebibliography}%

\end{document}